\documentclass[aps,pre,a4paper,twocolumn,superscriptaddress]{revtex4-1}
\pdfoutput=1

\usepackage{natbib} 
\usepackage{ucs}
\usepackage[utf8x]{inputenc}
\usepackage{amssymb,amsmath,bm}
\usepackage{graphicx}
\usepackage{eucal}
\usepackage{multirow}
\newcommand{\dd}{\mathrm{d}}
\newcommand{\mean}[1]{\langle #1\rangle}
\newcommand{\beq}{\begin{equation}}
\newcommand{\eeq}{\end{equation}}
\newcommand{\beqn}{\begin{eqnarray}}
\newcommand{\eeqn}{\end{eqnarray}}
\newcommand{\float}[2]{ {#1}\times 10^{#2}}

\newcommand{\XX}{\mathcal{X}}
\newcommand{\VV}{\mathcal{V}}

\begin{document}

\title{Rheology of sliding leaflets in coarse-grained DSPC lipid bilayers}
\author{Othmene Benazieb}
\affiliation{Institut Charles Sadron, CNRS and University of Strasbourg, 23 rue du Loess, F-67034 Strasbourg cedex 2, France}
\author{Claire Loison}
\affiliation{Institut Lumière Matière, UMR5306 Université Lyon 1-CNRS, Université de Lyon,
69622 Villeurbanne cedex, France}
\author{Fabrice Thalmann}
\affiliation{Institut Charles Sadron, CNRS and University of Strasbourg, 23 rue du Loess, F-67034 Strasbourg cedex 2, France}
\date{July 11 2020}
\begin{abstract}
  Amphiphilic lipid bilayers modify the friction properties of the surfaces on top of which they are deposited. In particular, the measured sliding friction coefficient is significantly reduced compared with the native surface. We investigate in this work the friction properties of a numerical coarse-grained model of DSPC (1,2-distearoyl-sn-glycero-3-phosphocholine) lipid bilayer subject to longitudinal shear. The interleaflet friction coefficient is obtained from out-of-equilibrium pulling and relaxation simulations. In particular, we gain access to the transient viscoelastic response of a sheared bilayer. The bilayer mechanical response is found to depend significantly on the membrane physical state, with evidence in favor of a linear response regime in the fluid but not in the gel region.
\end{abstract}

\maketitle

%%%%%%%%%%%%%%%%%%%%%%%%%%%%%%%%%%%%%%%%%%%%%%%%%%%%%%%%%%%%%%%%
%%% INTRODUCTION
%%%%%%%%%%%%%%%%%%%%%%%%%%%%%%%%%%%%%%%%%%%%%%%%%%%%%%%%%%%%%%%%

\section{Introduction}

%%%
\subsection{Rheological properties of phospholipid bilayers}

Glycerophospholipids are essential compounds of biological lipid bilayers. Their molecular structure comprises one bulky zwitterionic, hydrophilic headgroup (phosphatidylcholine) and two aliphatic hydrocarbon chains esterified around a glycerol molecule. With chain lengths comprising between 12 and 20 carbons atoms, these molecules self-assemble as flat bilayers made of two leaflets with tail to tail opposing lipid molecules. Common phospholipids (e.g. dipalmitoyl-phosphatidylcholine DPPC, distearoyl-phosphatidylcholine DSPC) do not interdigitate under standard conditions, and the leaflets are relatively weakly bound together~\cite{2010_Marsh}. Bilayer fluidity depends significantly on temperature~\cite{Cevc_Marsh_PhospholipidBilayers}. Moreover, most pure lipid systems encounters a sharp thermodynamic melting transition at a given temperature $T_m$ (41$^{\circ}$C for DPPC, 55$^{\circ}$C for DSPC)~\cite{Marsh_HandbookLipidBilayers2}. Above melting, lipid tails are isomerically disordered, weakly cohesive, conferring fluidity to the bilayer with Arrhenius dependence in temperature. Below melting, lipid tails adopt  all-trans conformations, are subject to stronger cohesion, displaying solid type dynamics at short time scales, while remaining a viscous fluid on longer scales. 

Coating a solid surface with a dense phospholipid monolayer modifies the sliding friction properties significantly. Experiments reports a significant decrease in the sliding friction coefficient when \textit{both surfaces} are covered with lipids in a dense, or \textit{gel} conformation~\cite{2006_Briscoe_Klein}. This issue is relevant in the field of biolubrication, such as for instance the mechanism of articular joints. As a matter of fact, synovial fluid combines lipid and biopolymer molecules for optimal lubrication, the role of each component being still a topic of investigation.

It is difficult to relate the macroscopic friction between a pair of surfaces with microscopic mechanisms involved at the molecular scale~\cite{Persson_SlidingFriction}. In the case of hydrated lipid bilayers, a lateral shear displacement involves the solvent viscosity, the sliding leaflet friction and possibly some sliding of the solvent on top of the hydrophilic bilayer surface. In the framework of linear response, sheared lipid bilayers display a viscous response, characterized by an \textit{interleaflet friction coefficient} $b$, a Newtonian transverse viscosity $\eta$ for the solvent, and a solvent-bilayer friction $b'$ which quantifies the importance of the sliding of the fluid at the bilayer interface. 

The experimental determination of $b$ (and $b'$) is difficult. Evans and Yeung suggested that $b$ dominates the resistance of a bilayer when pulling a lipid nanotube from a giant vesicle, with a micropipette or an optical tweezer device~\cite{1994_Evans_Yeung}. Tube pulling experiments have since become a standard protocol for probing membrane physical properties, including the case of living cells~\cite{2012_Gauthier_Sheetz}. Leroy \textit{et al.} were able to estimate the dissipation induced by the friction of the interfacial water beneath a supported lipid bilayer deposited onto a mica surface using a surface force apparatus (SFA)~\cite{2009_Leroy_Charlaix}. More recently, simulations by Schlaich \textit{et al.}~\cite{2017_Schlaich_Netz} investigated in details the nature of the friction between amphiphilic surfaces separated by a variable amount of interfacial water using atomistic molecular dynamics simulations. The competition between interleaflet and water layer frictions in stacks of sheared lipid bilayers was investigated in Bo\c{t}an \textit{et al.}~\cite{2015_Botan_Loison}. 
Seifer and Langer~\cite{1993_Seifert_Langer} showed how the relaxation dynamics of the transverse membrane undulation modes depend on $\eta$ and $b$, and interpreted in this way experimental data from inelastic neutron scattering~\cite{1993_Pfeiffer_Sackmann}. This formalism was successfully used by den~Otter and Shkulipa for estimating $b$ for various numerical model of lipids, using equilibrium molecular dynamics (MD)~\cite{2007_denOtter_Shkulipa}. M\"{u}ller and M\"{u}ller-Plathe showed how the bilayer friction and viscosity parameters could be obtained from reverse non-equilibrium molecular dynamics (RNEMD) simulations~\cite{2009_Muller_MullerPlathe}.
Falk \textit{et al.} managed to determine $b$ for a coarse-grained bilayer in both the fluid and gel states using RNEMD~\cite{2014_Falk_Loison} for shearing the solvent on both sides across the bilayer. In particular, the authors reached the conclusion that there was only minor sliding velocity effects at the solvent-lipid interface. Using a similar method, Zgorski et al. determined $b$ and the membrane transverse viscosity for DPPC Martini models~\cite{2019_Zgorski_Lyman}.   

The approaches of den Otter and Shkulipa, or Falk \textit{et al.} cannot easily be generalized to supported bilayers in close interaction with a flat solid surface. It is known for instance that a proximal solid surface influences the lipid diffusion dynamics, as shown in Scomparin \textit{et al.}~\cite{2009_Scomparin_Tinland}. There is therefore a need for simple approaches for determining the friction properties of lipid bilayers interacting with solid surfaces. 

A natural idea consist in pulling directly on various system components (lipid or water layers) and measuring the resulting velocity profile. Alternatively, information can be obtained by observing how a system initially prepared with mutual nonvanishing relative sliding velocities relaxes to its equilibrium state. When linear response from the system holds, it is expected on general grounds that both approaches give consistent results. In the present work we show how a constant pull force and momentum relaxation methods can be used to determine the interleaflet friction coefficient in the simple case of a coarse-grained lipid bilayer in water. 

%%%
\subsection{The Martini model}

Martini is a successful coarse-grained representation of lipids, with a 4~heavy atoms to 1~bead center level of coarse graining. This model displays a realistic fluid phase, as well as an ordered ``gel'' phase, with nematically oriented chains but disordered headgroups. In lipid biophysics, the gel phase corresponds to a viscous, almost solid, state of the lipids observed at low temperatures. The transition between gel and fluid phases is a weakly first order phase transition, called main or melting transition, accompanied by a discontinuous change in structural parameters such as the nematic ordering of the chains or the bilayer thickness. If the fluid phase is fairly well reproduced by the Martini model, which was designed for this purpose, the existence of a gel phase is a happy outcome of the model. While missing some characteristics of the experimental gel phase, the numerical low temperature phase captures some important features: stronger cohesion, larger thickness, lower molecular mobility. However, the Martini model misses the existence of a ripple phase P$_{\beta'}$ below the melting transition, and the presence of a chain tilt angle below the pretransition temperature $\mathrm{L}_{\beta'} \to~\mathrm{P}_{\beta'}$~\cite{Cevc_Marsh_PhospholipidBilayers,Heimburg_BiophysicsMembrane}. 

\begin{figure}[ht]
\centering
\includegraphics[width=0.46\textwidth,angle=0]{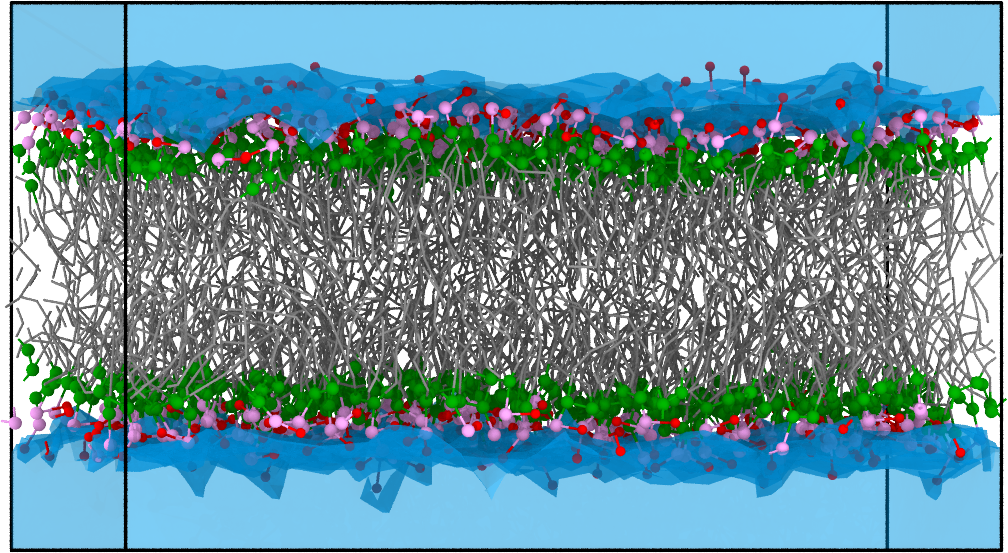}
\caption{Snapshot of a configuration of a coarse-grained bilayer containing 256 DSPC lipids per leaflet, with 2560 water beads molecules on both sides, in the high temperature fluid state at 340~K.}
\label{fig:Fig1_fluid_crop_snap}
\end{figure}

\begin{figure}[ht]
\centering
\includegraphics[width=0.46\textwidth,angle=0]{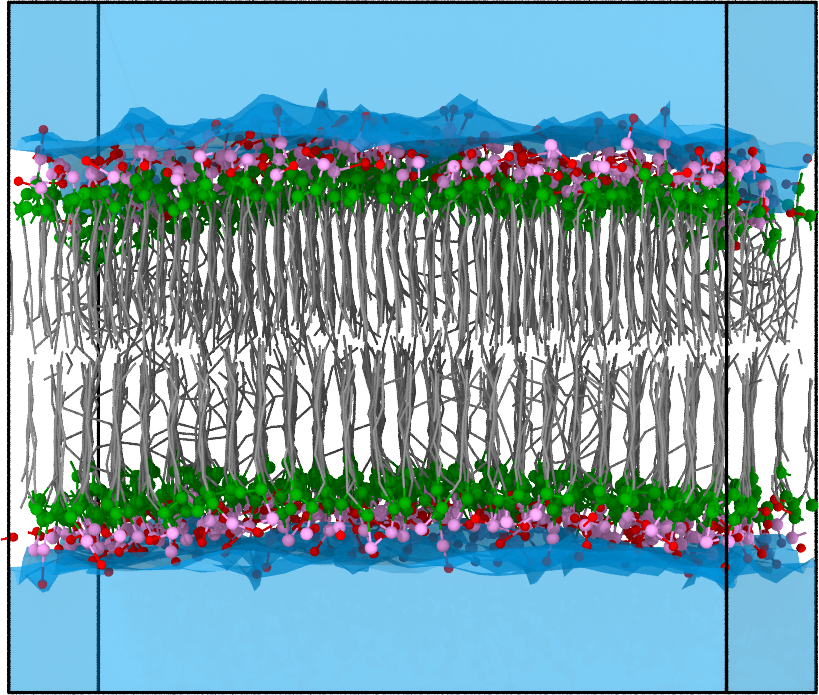}
\caption{Snapshot of a configuration in the low temperature gel state at 280~K. Compared to the fluid case, the bilayer is less extended in the $xy$ direction and thicker. No appreciable lipid chain tilt angle is visible.} 
\label{fig:Fig2_gel_crop_snap}
\end{figure}

We chose to study DSPC molecules, parameterized using the version~\textit{v2.0} of the Martini model~\cite{2007_Marrink_deVries}. DSPC lipids possess two saturated 18 carbons chains. This choice was driven by experimental considerations, as DSPC supported lipid bilayers obtained by Langmuir deposition constitute a robust and well studied model systems~\cite{2009_Scomparin_Tinland,2012_Hemmerle_Daillant} which we intend to simulate in a near future.  Our simulated systems comprise a single bilayer alongside a single water slab, with periodic boundary conditions in the three dimensions. Two representative snapshots are shown in Figs~\ref{fig:Fig1_fluid_crop_snap} and~\ref{fig:Fig2_gel_crop_snap}.

%%%%%%%%%%%%%%%%%%%%%%%%%%%
%%% Snapshot Fluid
%%%%%%%%%%%%%%%%%%%%%%%%%%%
%%%%%%%%%%%%%%%%%%%%%%%%%%%
%%%%%%%%%%%%%%%%%%%%%%%%%%%
%%% Snapshot Gel
%%%%%%%%%%%%%%%%%%%%%%%%%%%
%%%%%%%%%%%%%%%%%%%%%%%%%%%

The Martini model is designed in order to reproduce faithfully the structural and thermodynamic properties of lipids in the fluid phase~\cite{2004_Marrink_Mark,2007_Marrink_deVries}. Coarse-grained beads interaction potentials are not tabulated but assume a Lennard-Jones functional form, though with larger radii and energy parameters compared with the atomistic case. The standard Gromacs implementation~\cite{2008_Hess_Lindahl} of the Martini model uses standard molecular dynamics algorithms, such as Verlet integrator and Nose-Hoover or v-rescale weak coupling thermostats~\cite{Allen_Tildesley_Simulations_2nd, Frenkel_Smit_MolecularSimulations,2007_Bussi_Parrinello}. 
These design and implementation choices imply that the kinetic properties of the Martini systems do not quantitatively agree with atomistic simulations or experiments. The corresponding kinetic properties must therefore be discussed at a qualitative level, focusing on relative differences between situations, or investigating various methodological approaches.

%%%%%%%%%%%%%%%%%%%%%%%%%%%%%%%%%%%%%%%%%%%%%%%%%%%%%%%%%%%
%%% Methodology
%%%%%%%%%%%%%%%%%%%%%%%%%%%%%%%%%%%%%%%%%%%%%%%%%%%%%%%%%%%

\section{Methodology}

%%%%
\subsection{Relaxation and forced sheared experiments.}

Our purpose is to characterize the response of a supported bilayer sheared parallel to its longitudinal $xy$ directions, as it may provide clue on the experimentally observed drag friction reduction upon coating surfaces with deposited lipid mono or multilayers. The determination of the interleaflet friction bilayer in water solution is therefore a first step towards the desired answer, which will be later extended to lipid layers deposited onto solid surfaces.

Two strategies were used in the present work. Both were implemented using the Gromacs molecular dynamics simulation tool~\cite{2008_Hess_Lindahl}. In the first approach, here referred as \textit{constant pull force} (CPF), a non-equilibrium stationary pull of each membrane leaflets was set up, resulting in a constant drift velocity of the bilayer. The pulling force-velocity ratio gives access into the value of the interleaflet friction coefficient~$b$.  In the second approach, referred as \textit{force kick relaxation} (FKR), the relaxation stage of a leaflet consecutive to an initial step increase in its center of mass (COM) velocity was measured. The displacement response curve of the leaflet gives another estimate of the interleaflet coefficient $b$. It provides in addition a direct picture of the transient bilayer response following a sudden shear force kick. 

%In particular, we show that shear tilt modes of the bilayer are sollicitated, and that on the nanosecond (coarse-grained) time scale, the response is undoubtedly viscoelastic. The transient viscoelastic response is also present in the pulling force experiment, where it appears as a transient regime preceding the constant velocity regime.

A linear response regime is expected provided the pulling forces (CPF) and initial velocities (FKR) remain below their respective threshold values. In the linear regime, both drift velocities and displacements compete with random equilibrium fluctuations, a situation corresponding to a small Peclet number. The extraction of the signal (drift displacement and velocity) out of the noise (equilibrium fluctuations) requires averaging over many independent simulation runs. The statistical significance of the bilayer response curves was estimated by means of a bootstrap statistical procedure. In our case, for every simulation condition (external constant force, or initial force kick), a sample of \textit{ca} $N_s\sim 50, 150, 1000$ independent runs was subject to random reweighting, in order to infer a reliable value of the statistical uncertainty associated with sample averaging. Details on our numerical simulation procedure and the associated statistical analysis are deferred to the appendix section.

%%% 
\subsection{Standard hydrodynamic description}

\begin{figure}[ht]
    \centering
    \resizebox{0.46\textwidth}{!}{\includegraphics{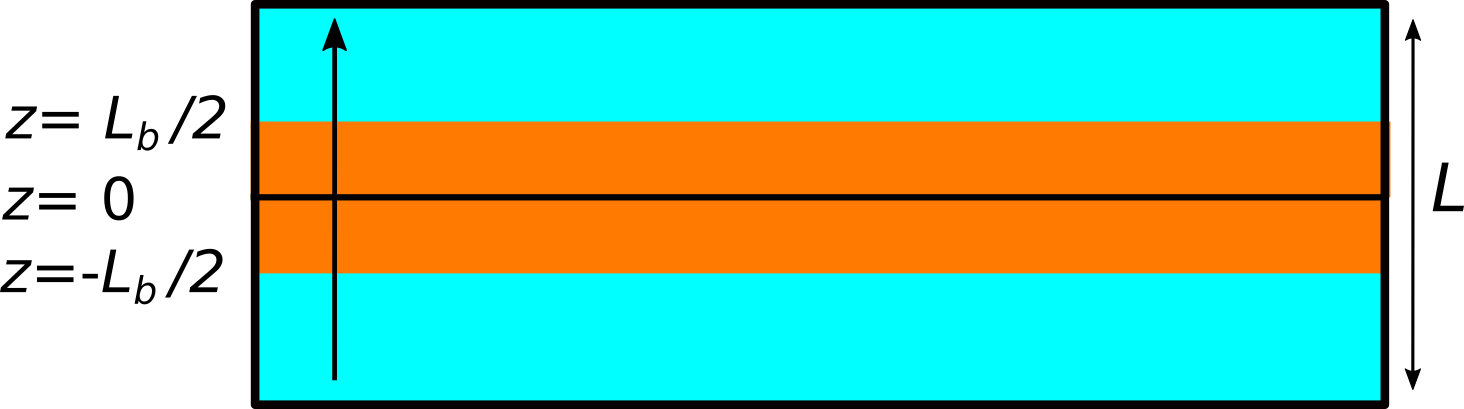}}
    \caption{Geometric parameterization of the system used in the present study, with $L=L_b+L_w$.}
    \label{fig:Fig3_slab}
\end{figure}

A natural interpretation frame for our numerical simulations is the classical hydrodynamics model. In this framework, both lipid leaflets are described as rigid solid slabs (thickness $L_b$, area $A$), surrounded by a water layer considered as a Newtonian fluid (thickness $L_w$, viscosity $\eta$). Inertia of lipids (leaflet mass $M$) and fluid (volumetric mass density $\rho$) components are accounted for. The upper and lower leaflets move with respective velocities $V_u,V_d$ along the horizontal $x$ direction. Water is described by a Eulerian velocity field $v(z)\vec{e}_x$, where the vertical coordinate $z$, normal to the bilayer, varies in the interval $z> L_b/2;\, z<-L_b/2$ with periodic boundary conditions $v(z+L)=v(z)$ (PBC), and $x$ is one of the horizontal direction, without loss of generality (Fig~\ref{fig:Fig3_slab}). The fluid is subject to a Newtonian shear stress $\tau_{zx}(z)$, abbreviated as $\tau(z)$. Sticking boundary conditions at the lipid water interface $z=\pm L_b/2$ are assumed (or equivalently an infinite lipid-fluid friction $b'=\infty$). 

%%%%%%%%%%%%%%%%%%%%%%%%%%%%%%%%%%%%%%%%%%%%%%%%%%%%
%%% Slab parameterization
%%%%%%%%%%%%%%%%%%%%%%%%%%%%%%%%%%%%%%%%%%%%%%%%%%%%
%%%%%%%%%%%%%%%%%%%%%%%%%%%%%%%%%%%%%%%%%%%%%%%%%%%%

%Due to periodicity, leaflets mutually interact at $z=\pm L_w/2+L$. 

We assume that leaflets experience a friction proportional to their mutual relative sliding velocity $V_u-V_d$, leading to an interleaflet shear stress $\tau=\tau(z=0)$ obeying
\beq
\tau = b (V_u-V_d),
\eeq
with $b$ the interlayer friction coefficient. An average fluid velocity can be defined as:
\beq
V_w = \frac{1}{L_w}\int_{L_b/2}^{L_b/2+L_w\equiv -L_b/2 [L]} \dd z\, v(z).
\eeq
In addition, we consider the possibility to act upon each leaflet, and the water layer, by means of a uniform force acting on the center of mass of the corresponding subsystem. Such forces are respectively denoted $F_u, F_d, F_w$, and directed along $x$. For convenience, one introduces the corresponding stresses $\phi_{\mu}$ ($F_{\mu}=A\phi_{\mu}$) with $\mu=u$ (upper leaflet), $\mu=d$ (lower leaflet) and $\mu=w$ (water region). One restricts ourselves to the physical case of a vanishing total force $F_u+F_d+F_w=0$, henceforth preserving the total momentum $\rho L_w V_w + MV_u + MV_d$ of the hydrodynamic system.

The stationary solution of the hydrodynamic problem corresponds to a parabolic flow. Two stationary velocity profiles are of particular interest. The linear Couette profile corresponds to $\phi_u= -\phi_d$, $\phi_w=0$, $V_u=-V_d$, $V_w=0$ and
\beq
2\left(b+\frac{\eta}{L_w}\right)V_u=\phi_u.
\label{eq:stationaryCouetteFlow}
\eeq
The Poiseuille flow profile corresponds to $\phi_u=\phi_d = -\phi_w/2$, $V_u=V_d$ and
\beq
6 \frac{\eta}{L_w}(V_u-V_w)=\phi_u = -\frac{\phi_w}{2}.
\label{eq:PoiseuilleStationary}
\eeq
The relation above can be further simplified as the total momentum is assumed to vanish $2M \phi_u +\rho L_w \phi_w=0$. Both flows are represented in Fig~\ref{fig:Fig4_Couette_Poiseuille_flow}.\medskip

% MARTINI WATER viscosity  0.79~mPs.
% Careful of temperature and pressure conditions

\subsection{Viscoelastic relaxation model.}

As the Results section demonstrates, the hydrodynamic model is useful but does not accurately represent the observed numerical behavior. We therefore introduce here a more general viscoelastic model.  We assume that a transient linear response of a bilayer subject to a suddenly applied external force exists, that can be expressed by means of a retarded memory function. Using the same notations as above, but now with time dependent velocity fields $V_u(t),V_d(t),V_w(t)$ one has:
\begin{widetext}
\beqn
\frac{M}{A}\dot{V}_u(t) &=& \int_{-\infty}^{t} \dd s\, \left[g_{\mathrm{bu}}(t-s) (V_d(s)-V_u(s)) + g_{\mathrm{wu}}(t-s) (V_w(s)-V_u(s))\right] +\phi_u(t);\\
\frac{M}{A}\dot{V}_d(t) &=& \int_{-\infty}^{t} \dd s\, \left[g_{\mathrm{bu}}(t-s) (V_u(s)-V_d(s)) + g_{\mathrm{wu}}(t-s) (V_w(s)-V_d(s))\right] +\phi_d(t);\\
\rho L_w\dot{V}_w(t) &=& \int_{-\infty}^{t} \dd s\, \left[g_{\mathrm{wu}}(t-s) (V_u(s)+V_d(s)-2V_w(s))\right] +\phi_w(t).
\eeqn
\end{widetext}
The retarded response involves two memory functions. A first kernel $g_{\mathrm{bu}}(t)$ accounts for the interleaflet interaction, including interleaflet dynamic friction, lipid inertia as well as viscoelastic lipid elastic tilt and stretch modes. A second kernel $g_{\mathrm{wu}}(t)$ accounts for all the water leaflet interactions, which possibly includes solvent sliding friction, retardation of the fluid motion due to inertia, and again viscoelasticity arising from lipid tilt and stretch. The same kernel is used for both leaflets, as a consequence of the up-down $z$ symmetry of the flow. External stresses $\phi_u(t),\phi_d(t),\phi_w(t)$ are arbitrary functions of time.

\begin{figure}[ht]
\centering
\includegraphics[width=0.46\textwidth,angle=0]{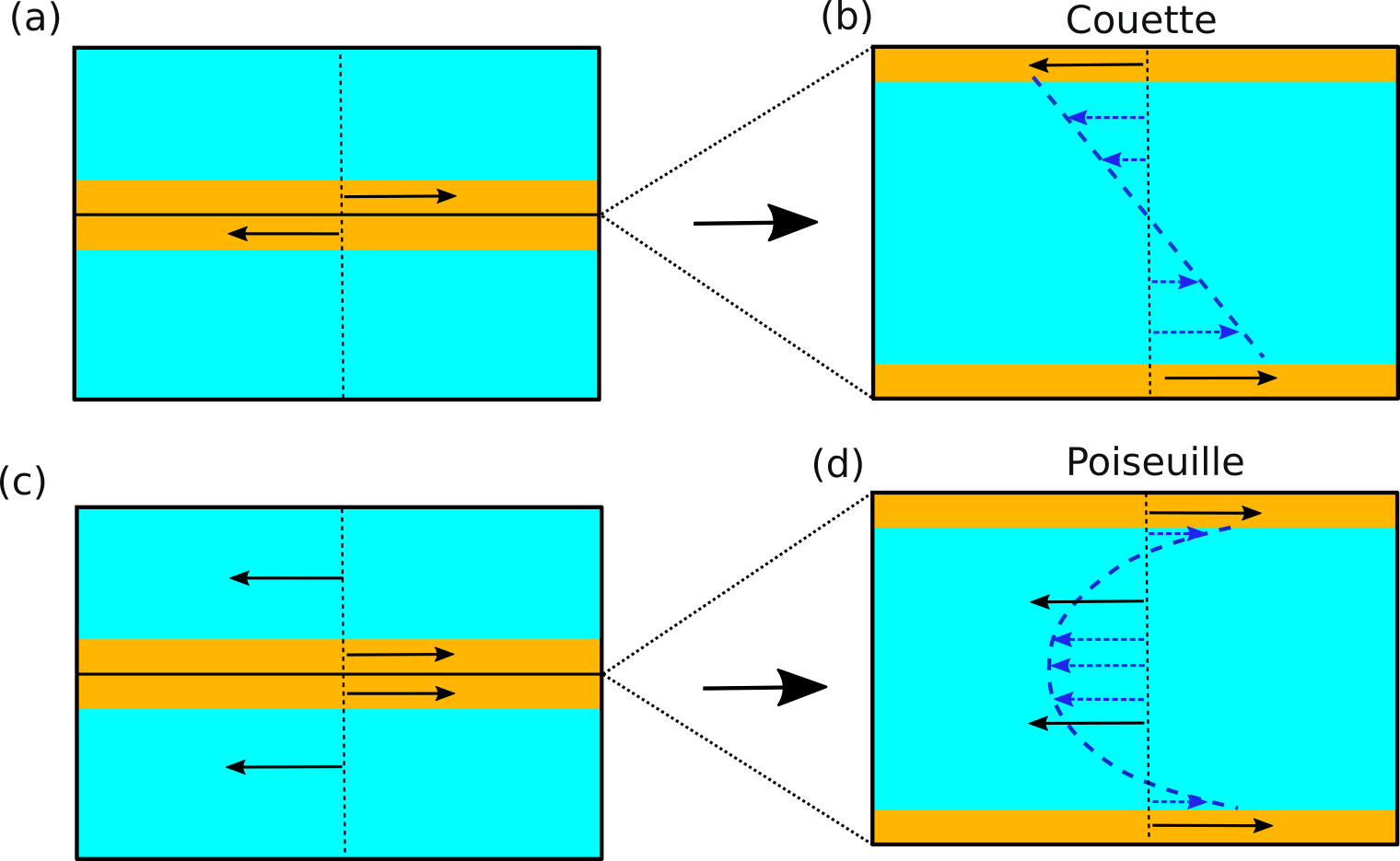}
\caption{When opposing forces are exerted on each leaflet (case a) the resulting stationary state sees the two leaflets sliding at constant relative velocity, surrounded by a uniform solvent velocity gradient profile, as emphasized in the sub-picture (b) where the simulation box boundary has been purposely shifted to sit exactly at the mid-plane of the bilayer. Case (a-b) is subsequently referred as a \textit{Couette situation}. When a uniform force is exerted on both leaflets and an opposing force on the solvent beads (case c), a symmetric parabolic velocity profile builds up in the solvent, assuming sticking boundary conditions at the interface with the bilayer (d). Case (c-d) is subsequently referred as \textit{Poiseuille situation}. In all cases, the total momentum of the system is constant and vanishes.}
\label{fig:Fig4_Couette_Poiseuille_flow}
\end{figure}

We now restrict ourselves to two main situation of interests, namely Couette $V_w=0$, $V_u(t)=-V_d(t)$, $\phi_u(t)=-\phi_d(t)$, $\phi_w=0$ and Poiseuille $V_u(t)=V_d(t)= -V_w(t)\rho L_w/2M$, $\phi_u(t)=\phi_d(t)=-\phi_w(t)/2$ (see Fig~\ref{fig:Fig4_Couette_Poiseuille_flow}). The retarded motion equations are in the Couette case:
\begin{eqnarray}
\frac{M}{A}\dot{V}_u &=& -\int_{\infty}^{t} \dd s\, (2g_{\mathrm{ud}}+g_{\mathrm{wu}})(t-s) V_u(s) + \phi_u(t);\nonumber\\
V_w &=&0,
\label{eq:retardedCouette}
\end{eqnarray}
and in the Poiseuille case:
\begin{eqnarray}
\frac{M}{A}\dot{V}_u &=& -\int_{\infty}^{t} \dd s\, g_{\mathrm{wu}}(t-s)\left(1+\frac{2M}{A\rho_w L_w}\right) V_u(s)\nonumber\\
& & + \phi_u(t);\nonumber\\
\rho L_w \dot{V}_w &=& -\int_{\infty}^{t} \dd s\, g_{\mathrm{wu}}(t-s)\left(2+\frac{A\rho_w L_w}{M}\right) V_w(s)\nonumber\\
& &+ \phi_w(t).
\label{eq:retardedPoiseuille}
\end{eqnarray}

Of particular importance in the present study is the response to a couple of force kicks (Couette case)
\beq
\phi_u= -\phi_d= \frac{M}{A}V_0 \delta(t),
\eeq
that confers instantly a momentum $MV_0\vec{e}_x$ to the upper leaflet, and $-MV_0\vec{e}_x$ to the lower leaflet. Velocity profiles can be inversed by Laplace transforms of the velocity, stress and memory functions, \textit{e.g.}
\beq \hat{V}_u(p) = \int_0^{\infty}\dd t\, e^{-pt} V_u(t),\eeq
leading to
\beq
\left( \frac{M}{A} p + 2\hat{g}_{\mathrm{ud}}+\hat{g}_{\mathrm{wu}}\right) \hat{V}_u(p) = \frac{M}{A} V_u(t=0).
\eeq

In particular, the impulsional displacement $\Delta X_u= \int_0^{\infty}\dd t\, V_u(t) = \hat{V}(p=0)$ obeys the relation
\beq
\Delta X_u = \frac{\displaystyle \vphantom{\int}\frac{M}{A}V_u(0)}{\vphantom{\int}2\hat{g}_{\mathrm{ud}}(0)+\hat{g}_{\mathrm{wu}}(0)}.
\eeq
In the mean time, a stationary stress $\phi_u$ should result in an asymptotically constant velocity $V_u$ and $\dot{V}_u = 0$:
\begin{eqnarray}
\int_0^{\infty} \dd s\,(2g_{\mathrm{ud}}(t)+g_{\mathrm{wu}}(t)) V_u &=& (2\hat{g}_{\mathrm{ud}}(0)+\hat{g}_{\mathrm{wu}}(0)) V_u \nonumber\\
&=& \phi_u
\end{eqnarray}
One recovers the hydrodynamic limit $\phi_u/V_u = 2(b+\eta/L_w)$ and therefore
\beq
(2\hat{g}_{\mathrm{ud}}+\hat{g}_{\mathrm{wu}})(p=0) = 2\left(b+\frac{\eta}{L_w}\right)
\eeq

In conclusion, one obtains a useful relation between the impulsional displacement $\Delta X_u$ and the interleaflet friction coefficient.

\beq
b+\frac{\eta}{L_w} = \frac{\displaystyle \frac{M}{A}V_u(0)\vphantom{\int}}{2\Delta X_u\vphantom{\int}}
\label{eq:ImpulsionFrictionCoefficient}
\eeq

This viscoelastic model assumes a linear relation between forces (the cause) and displacement or velocity (the effect). A master curve $\Xi_{(\phi)}(t)$ can be introduced to represent the normalized drift displacement $(X_u(t)-X_u(0))/\phi_u$ associated to a step stress $H(t)=1$ for $t\geq 0$ and $H(t)=0$ for $t<0$ (Heaviside function).  This master curve obeys
\begin{eqnarray}
\Xi_{(\phi)}(t)&=& 0\; \mathrm{for}\; t<0; \\
\frac{M}{A}\ddot{\Xi}_{(\phi)}(t) &=& -\int_{\infty}^{t} \dd s\, (2g_{\mathrm{ud}}+g_{\mathrm{wu}})(t-s) \dot{\Xi}_{(\phi)}(s)\nonumber\\
& & + H(t) \; \mathrm{for}\; t\geq 0.
\label{eq:masterCurveDisplacement}
\end{eqnarray}
In the mean time, a master curve for the normalized displacement $\Xi_{(V)}(t)= (X_u(t)-X_u(0))/V_0$ can be introduced for the impulsion case, which obeys:
\begin{eqnarray}
\Xi_{(V)}(t)&=& 0\; \mathrm{for}\; t<0; \\
\frac{M}{A}\ddot{\Xi}_{(V)}(t) &=& -\int_{\infty}^{t} \dd s\, (2g_{\mathrm{ud}}+g_{\mathrm{wu}})(t-s) \dot{\Xi}_{(V)}(s) \nonumber\\
& & + \frac{M}{A}\delta(t)\; \mathrm{for}\; t\geq 0.
\label{eq:masterCurveVelocity}
\end{eqnarray}
Both master curves can be related to the memory function $2g_{\mathrm{ud}}+g_{\mathrm{wu}}$ in Laplace space.

%%%%%%%%%%%%%%%%%%%%%%%%%%%
%%% Couette and Poiseuille
%%%%%%%%%%%%%%%%%%%%%%%%%%%
%%%%%%%%%%%%%%%%%%%%%%%%%%%

%%%
\subsection{Diffusion of the lipids and water centers of mass}

Simulations deal with finite size systems, and thermal fluctuations are always present. In our case, the center of mass of each of the three main components of the simulated system (upper and lower leaflets, water) is subject to brownian motion, while the global center of mass is fixed, as required by weak coupling to Nose-Hoover or v-rescale thermostats. It results that the instantaneous kinetic energy of the upper, lower leaflets and water is not given by the usual equipartition of energy theorem. However, the order of magnitude of the instantaneous kinetic energies $M_{\mu}V_{\mu}^2/2, \mu=\{h,u,w\}$ remains of the order of $k_B T/2$. 

We therefore distinguish the average, non fluctuating hydrodynamic displacements $X_{\mu}(t), \mu = \{h,u,w\}$ from the sampled, brownian trajectories $\XX_{\mu}^{(\alpha)}(t)$, with $\alpha$ an index relative to a given center of mass trajectory realization, or simply $\XX_{\mu}(t), \mu = \{h,u,w\}$ when referring to a generic trajectory. Similarly, one introduces the brownian instance of the velocity response $\VV_{\mu}^{(\alpha)}(t)$ or generically $\VV_{\mu}(t)$.

In order to quantify the magnitude of the brownian fluctuations acting on the positions $\XX_{\mu}(t)$, one naturally defines the diffusion coefficient $D_{\mathrm{COM,\mu}}$ of the center of mass of the subcomponent $\mu$ (not to be confused with the molecular diffusion coefficient), based on the mean quadratic displacements $\mean{(\XX_{\mu}(t)-\XX_{\mu}(0))^2}$. Hydrodynamic and brownian displacements are related by canonical ensemble averages $X_{\mu}(t)=\mean{\XX_{\mu}(t)}$. So are the velocities  $V_{\mu}(t)=\mean{\VV_{\mu}(t)}$.

Expression~(\ref{eq:ImpulsionFrictionCoefficient}) relates the dissipation  $b+\frac{\eta}{L_w}$ to the normalized displacement $\Delta X_u/V_u(0)$.
Noting that $\Delta X_u/V_u(0) = \int_0^{\infty} \dd t\, V_u(t)/V_u(0)$, one can write
\begin{eqnarray}
\frac{A\Delta X_u}{M V_u(0)} &=& \frac{A}{M V_u(0)^2} \int_0^{\infty} \dd t\, V_u(t)V_u(0)\nonumber\\
& = & \frac{2}{b+\eta/L_w}.
\label{eq:diffusionDissipation}
\end{eqnarray}
By analogy with brownian motion, where the diffusion coefficient is linked to the velocity autocorrelation function, one has $2D_{\mathrm{COM,u}}t \simeq 2 t\int_0^{\infty} \dd t\, \mean{\VV_u(t)\VV_u(0)}$, and obtain from (\ref{eq:diffusionDissipation}) a heuristic "Stokes-Einstein" relation:
\beq
D_{\mathrm{COM,u}} \sim \frac{2\mean{M \VV(0)^2}}{A(b+\eta/L_w)} \sim \frac{k_BT}{A(b+\eta/L_w)}.
\label{eq:heuristicEinsteinRelation}
\eeq
The precise relation between the relative quadratic displacements matrix of the various system subcomponents (leaflets, water\ldots) and the hydrodynamic friction coefficients ($b,\eta$\ldots) when the global center of mass is fixed is non trivial and will be the subject of future work. Eq.~\ref{eq:heuristicEinsteinRelation} provides however an order of magnitude for $D_{\mathrm{COM,u}}$.

%%%
\subsection{Constant pulling force simulations.}

A direct estimate of the asymptotic stationary drift velocity $\mean{\VV_{\mu}(t)}$  obtained as a result of a piecewise constant step increase of the external applied stresses $\phi_u=-\phi_d, \phi_w=0$ can be obtained by pulling directly on the leaflets. Even though the out of equilibrium features of the molecular dynamics software that we use are somewhat limited, it is possible to exert a constant force to the upper leaflet while exerting the opposite force on the lower leaflet~(section~\ref{app:SimulationDetails}). This features comes as part of tools available to perform biased, constrained umbrella sampling simulation schemes. The displacement $\XX_u(t)$ can then be read directly from the trajectory and its average value $\mean{\mathcal{X}_u}$ fitted to an affine time function $x_0+tV_u$.

The possibility of imposing a pulling force for long times enables a quite precise determination of the relative stationary drift velocity of the leaflets. 

%%%
\subsection{Force kick relaxation simulations.}

Starting from an equilibrium trajectory configuration (reference NVT run), an initial condition $\mathcal{C}_{\alpha}$ is prepared by adding an identical $V_0\vec{e}_x$ constant velocity to all the beads pertaining to the upper leaflet, and the opposite velocity to all the beads in the lower leaflet. In the Martini model, all beads possess the same mass (72 a.m.u., 1008~Da for a DSPC molecule), and the upper leaflet center of mass acquires a finite momentum $M V_0\vec{e}_x$ as a result, with $M$ the mass of all beads in a leaflet. The velocity of the water beads is unaltered.
Physically, this corresponds to an instantaneous force torque $(M V_u\vec{e}_x\delta(t), -M V_u\vec{e}_x\delta(t))$ applied to the bilayer, and the total momentum of the system is preserved. In particular, the system center of mass remains fixed, as required when using a Nose-Hoover or velocity-rescale thermostat. Following the force kick, the kinetic energy of the bilayer is increased by an amount
\begin{eqnarray}
\sum_{i=1}^{N_b} \frac{m}{2} (\vec{v}_i\pm V_0\vec{e}_x)^2 &=& \sum_{i=1}^{N_b} \frac{m}{2} \vec{v}_i^2 + \frac{N_b m}{2}V_0^2\nonumber\\
& & + V_0\cdot(\sum_{i=1}^{N_b} \pm \vec{v}_i.\vec{e}_x),
\label{eq:InitialVelocityShift}
\end{eqnarray}
where $N_b$ stands for the number of beads (center of forces) present in the moving leaflet, and $m$ the associated (here identical) bead masses. The third term is a statistical $\mathcal{O}(\sqrt{N_b})$ fluctuation. The kinetic energy term is therefore increased by a relative amount
\beq
\frac{M V_0^2}{3N_b k_b T}.
\label{eq:MaximalVelocity}
\eeq
This sets an upper bound $V_{\mathrm{max}}$ for the velocity shift $V_0$ that can be applied without requiring the thermostat to pump too much energy out of the system, of the order of $V_{\mathrm{max}}= (N_b k_b T/M)^{1/2}\simeq 0.2~\mathrm{nm.ps}^{-1}$, using $N_b=256\times 14= 3584$, $M = N_b\times 72~\mathrm{amu}$ and $T=340$~K. 

Assigning to each leaflet a too small initial velocity value results in lowering the signal to noise ratio, the signal being the forward displacement and the noise the brownian displacement of the leaflet center of mass. Assuming  it takes a characteristic time $t_{\mathrm{relax}}$ for the leaflets to return to equilibrium, and that a given initial drift velocity $V_0$ drives the leaflet over a distance $\Delta X_u$, the ratio between ballistic and random displacement reads $\Delta X_u/\sqrt{D_{\mathrm{COM,u}} t_{\mathrm{relax}} }$ at the end of the relaxation stage. 
If in addition, the simple and naive scaling $\Delta X_u=V_0 t_{\mathrm{relax}}$ holds, the ballistic to random displacement ratio assumes a familiar Peclet number expression $\mathrm{Pe}^{1/2}$ with $\mathrm{Pe}=V_0\Delta X_u/D_{\mathrm{COM,u}}$. 

The  displacement $\Delta \XX_u(t)$ is monitored as a function of time $t$. As each run provides a noisy  brownian response $\Delta \XX^{(\alpha)}_u(t)$, the procedure must be repeated many times, until a significant displacement $\mean{\Delta \XX_u(t=\infty)}$ emerges from the thermal noise. Meaningful information can only be obtained in the linear response regime, \textit{i.e.} when the ratio $\Delta X_u/V_u(0)$ is constant up to some uncertainty. Too large velocity kicks $V_u(0)\gg V_{\mathrm{lr}}$ deviate from the linear regime and cannot be described within the framework of retarded linear response functions. The velocity scale $V_{\mathrm{lr}}$ until which the linear regime is expected to hold must be empirically determined and is expected to be smaller than $V_{\mathrm{max}}$ determined above. In the opposite limit, a too low kick $V_0$ does not give any useful result as the signal to noise ratio becomes too large. Again, to estimate a confidence interval for $\mean{\Delta \XX_u}$, one resorts to a statistical bootstrap procedure.

%%%
\subsection{Bootstrap procedure.}

The bootstrap is an empirical statistical method that provides a quantitative estimate for the confidence interval of an average sampled quantity~\cite{Press_Flannery_NumericalRecipes_C}. In the absence of extra information regarding the nature of the statistical process under investigation, the bootstrap approach uses only available sample values to build this estimate.

Considering a set of $N_s$ independent sampled values $\mathcal{S}_{0} = \{ x^{(\alpha)}\}, \alpha= 1\ldots N_s$ as main input information, one can generate an number $M$ of synthetic samples $\mathcal{S}_{\beta} = \{ x_{\beta}^{(\alpha)} \}, \alpha=1\ldots N_s$, $\beta = 1\ldots M$ by drawing with repetition, at random, $N_s$ elements of $\mathcal{S}_{0}$. The variability of the average
\beq
\langle f \rangle_{\beta} = \frac{1}{M} \sum_{\alpha=1}^{N_s} f(x_{\beta}^{(\alpha)})
\eeq
as a function of the synthetic samples $\mathcal{S}_{\beta}$, provides us with a confidence interval $2\sigma_b$ for the sampled average, using the following estimator
\beq
\sigma_b^2 \equiv  \frac{1}{M-1}\sum_{\beta=1}^{M}\left(\langle f \rangle_{\beta}-\frac{1}{M}(\sum_{\beta'} \langle f \rangle_{\beta'})\right)^2
\label{eq:bootstrapVariance}
\eeq
with $M$ large enough. In our case $M$ varies between 10 and 500. As discussed in~\cite{Press_Flannery_NumericalRecipes_C}, the bootstrap approach makes optimal use of the sole available information contained in $\mathcal{S}_{0}$.

%%%
\subsection{Preparation of the initial configurations.}
       
The system was equilibrated first at 340~K (fluid phase) and 280~K (gel phase) using a thermostat and a semi-isotropic barostat (see section~\ref{app:SimulationDetails}). This thermalization stage makes it possible to determine the average system size in the absence of external stress, or equivalently vanishing surface tension, respectively in the fluid and the gel phases. Out of equilibrium simulations were then run a number of times, using a thermostat and constant box size conditions $(L_x,L_z)$, where $L_x,L_z$ were the result of the previous step. Coupling to a thermostat was however still required to preserve the mechanical energy of the system.  For each phase, configurations from a reference canonical, constant volume (NVT) runs were then periodically recorded and stored, providing a set of up to 1000 initial conditions, in relation with the bootstrap and ensemble averaging procedures. The resulting equilibrium lipid bilayer geometrical characteristics are summarized in Table~\ref{table:GeometryBilayer}. 

%%%%%%%%%%%%%%%%%%%%%%%%%%%%%%%%%%%%%%%%%%%%%%%%%%%%%%
\begin{table}[]
    \centering
    \begin{tabular}{|c|c|c|c|c|c|}
    \hline
         State & $L_x$ (nm) & $L_z$ (nm) & $A$ (nm$^2$) &  $L_b$ (nm)  & $L_w$ (nm) \\
         \hline
         Fluid & 13.2 & 8.2 & 174. & 4.6 & 3.6 \\
         \hline
         Gel & 11.1 & 10.6 & 124. & 5.6 & 5.1 \\
    \hline
    \end{tabular}
    \caption{Geometric characteristics of the simulated systems in the fluid and gel regimes.}
    \label{table:GeometryBilayer}
\end{table}
%%%%%%%%%%%%%%%%%%%%%%%%%%%%%%%%%%%%%%%%%%%%%%%%%%%%%%

%%%%%%%%%%%%%%%%%%%%%%%%%%%%%%%%%%%%%%%%%%%%%%%%%%%%%%
%%% Results
%%%%%%%%%%%%%%%%%%%%%%%%%%%%%%%%%%%%%%%%%%%%%%%%%%%%%%
\section{Results}

%%%%%%%%%%%%%%%%%%%%%%%%%%%%%%%%%%%%%%%%%%%%%%%%%%%%%%
\subsection{Fluid phase constant pull force (CPF) simulations}

\begin{widetext}

%%%%%%%%%%%%%%%%%%%%%%%%%%%
%%% Trajectory, average and bootstrap (fluid)
%%%%%%%%%%%%%%%%%%%%%%%%%%%
\begin{figure*}[]
\centering
\includegraphics[width=0.9\textwidth,angle=0]{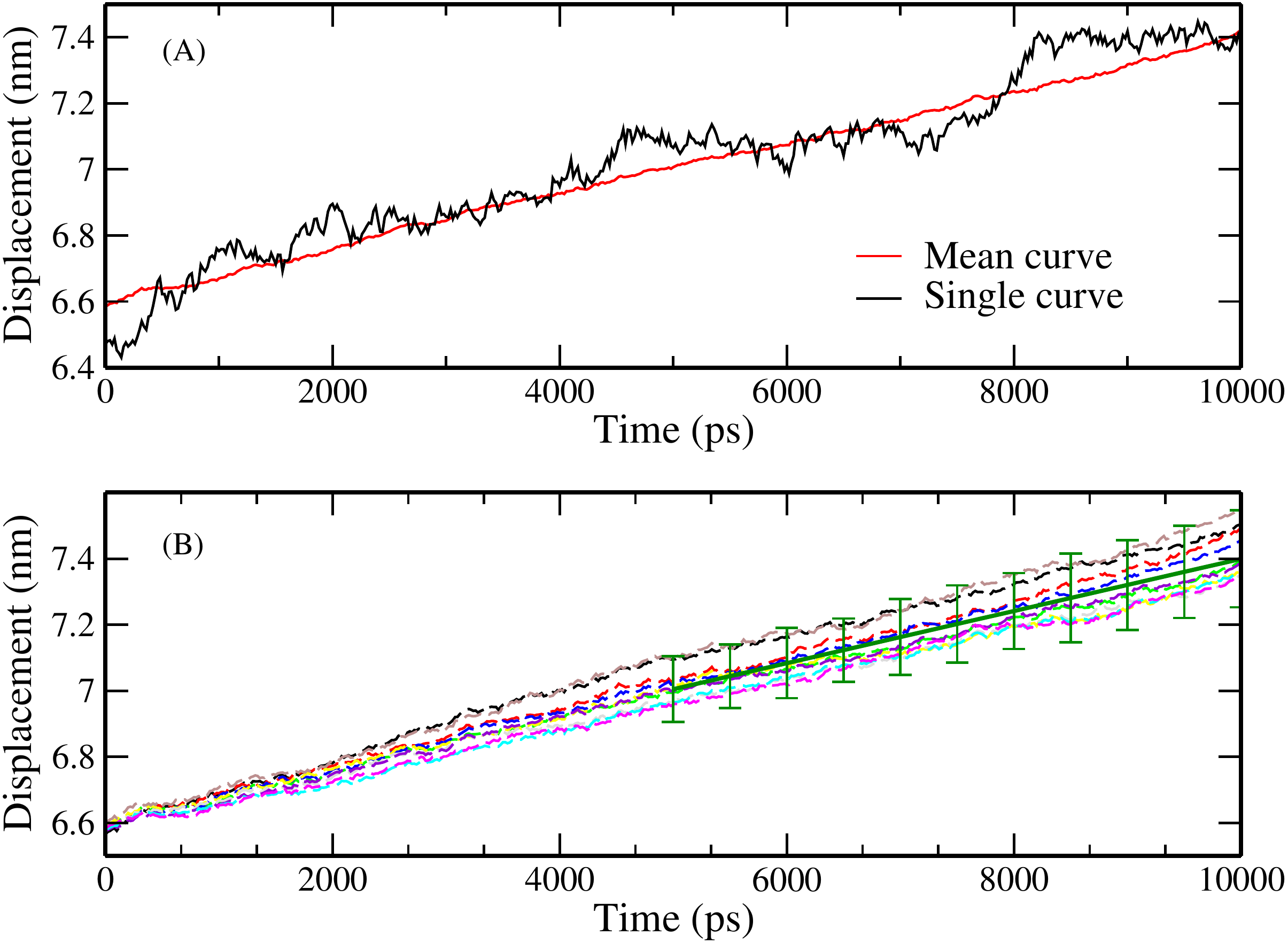}
\caption{Constant force pulling experiments in the fluid state: (A) single leaflet COM displacement $\mean{\XX}(t)^{(\alpha)}$, starting from the simulation box center (ca 6.6 nm) and averaged displacement $\mean{\XX}(t)\simeq 1/50 \sum_{\alpha=1}^{50} \XX^{(\alpha)}(t)$  \textit{vs} time. A bootstrap procedure (B) estimates the dispersion $\sigma_b(\XX_u(t))$ caused by the finiteness of the sample $\{\alpha\}$. Vertical bars represent the confidence interval of 10 selected points from the second half of the trajectory ($5000 < t < 10000$ ps), taken as twice the estimated bootstrap standard deviation. The vertical bars are used to provide a confidence interval for the drift velocity (slope of the averaged displacement curve).}
\label{fig:Fig5_illustration_displacement_leaflet}
\end{figure*}
%%%%%%%%%%%%%%%%%%%%%%%%%%%
\end{widetext}

The bilayer was submitted to a sequence of increasing pulling stresses $\phi_u$, resulting in an average displacement curve $\mean{\XX_u}$. Each external pulling force condition  was repeated about 50 times (Table \ref{table:ConstantStressRuns}), resulting in a sample set of raw displacement curves Fig~\ref{fig:Fig5_illustration_displacement_leaflet}(A). As seen in this figure, a typical pulling experiment generates a brownian displacement of the leaflet center of mass superimposed with a constant velocity horizontal translation. Panel~(A) superimposes  a raw displacement with an average over 50 equivalent displacements.  An example of bootstrap averaging of the trajectories is shown in Fig~\ref{fig:Fig5_illustration_displacement_leaflet}(B). Displacements curves start with a short transient regime, dominated by inertial and viscoelastic contributions. It is followed by a linear regime associated with stationary hydrodynamic dissipation and constant velocity translation $V_u$. The bootstrap analysis shows a dispersion among synthetic displacement curves, only slowly decreasing with the size of the set of trajectories, and inversely proportional to the applied stress $\phi_u$. 

\begin{figure}[ht]
\centering
\includegraphics[width=0.48\textwidth,angle=0]{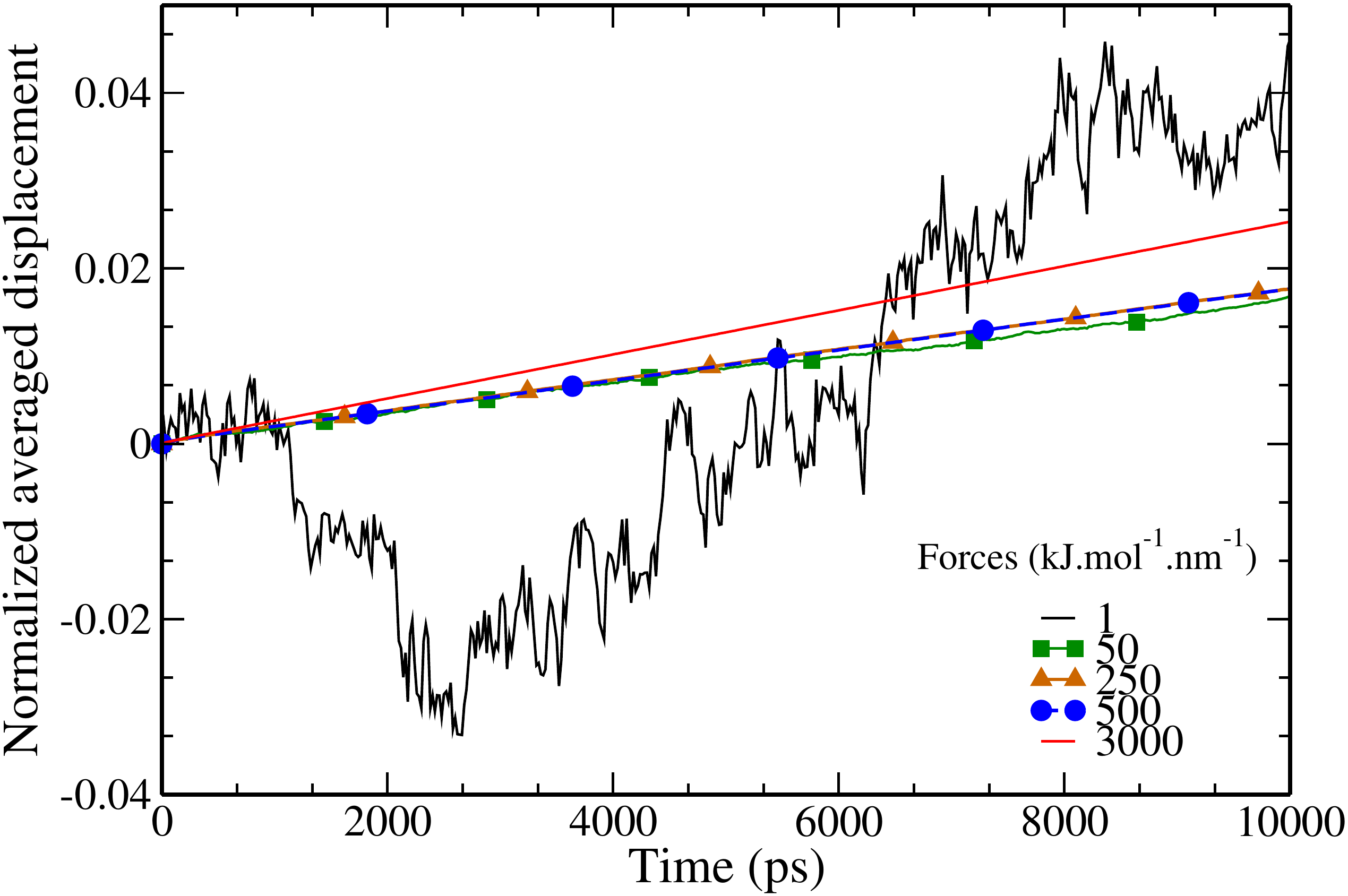}
\caption{Normalized averaged displacements (upper leaflet) $\mean{\XX}_u(t)/F$ for a set of increasing pulling forces $10,\ldots,3000$ kJ.mol$^{-1}$.nm$^{-1}$ (equivalently stresses $\tau\simeq 1\ldots 300$ bars).  The displacement for $F=3000$ lies clearly beyond the linear regime and the force $F=10$ competes with thermal agitation.}
\label{fig:Fig6_search_linear_regime}
\end{figure}

\begin{figure}[ht]
\centering
\includegraphics[width=0.48\textwidth,angle=0]{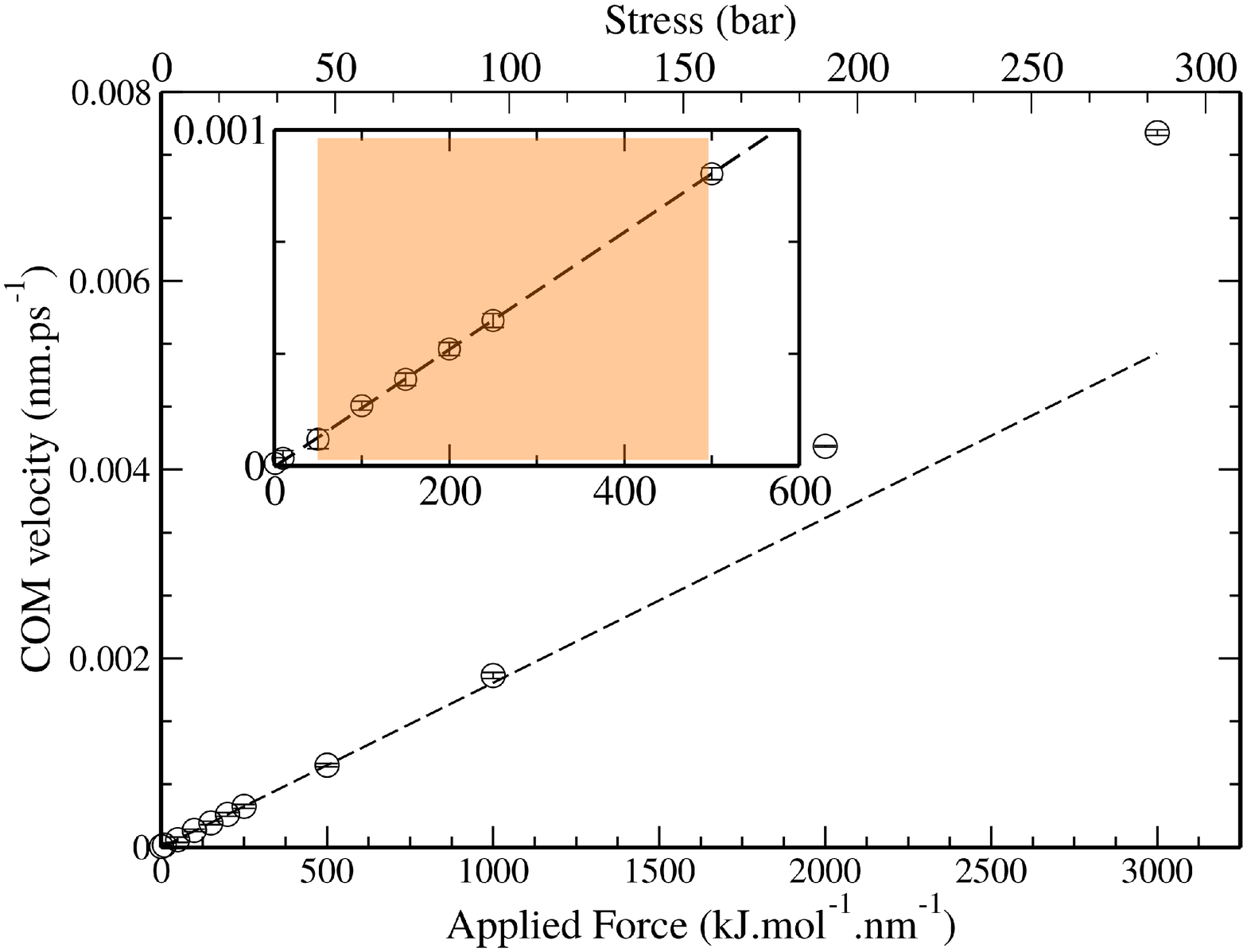}
\caption{Average drift velocities \textit{vs} applied forces (lower horizontal scale bar) or stresses (higher horizontal scale bar). A shear thinning deviation is seen at $\tau \geq 50$ bars. Inset: focus on the linear regime region. }
\label{fig:Fig7_linear_regime_pull_fluid}
\end{figure}

Averages of the normalized displacement curves  $\mean{\XX_u(t)}/F_u$ are shown in Fig~\ref{fig:Fig6_search_linear_regime}. In the framework of linear response, the averaged normalized displacements are expected to converge to a master curve $\Xi_{(\phi)}(t)$. This is indeed the case for a set of applied stresses within an interval $\float{4.8}{5}\leq \phi_u\leq \float{48}{5}$~Pa (applied forces in the range $50\leq F\leq 500~\mathrm{kJ.mol}^{-1}.\mathrm{nm}^{-1}$). A too small applied stress  $\phi_u=10^5$~Pa (force $F=10~\mathrm{kJ.mol}^{-1}.\mathrm{nm}^{-1}$) departs from the master curve due to strong brownian fluctuations\footnote{In the present case, a value $A=174~\mathrm{nm}^2$ was used for the area in the force-stress conversion, with  $\phi_u = 9600 F$, $\phi_u$ in Pa and $F$ in $\mathrm{kJ.mol}^{-1}.\mathrm{nm}^{-1}$. The conversion scale is also $\phi_u=0.096F\simeq 0.1 F$ if one whishes to express $\phi_u$ in bars.}. Large applied stresses clearly bring about strong deviations from linear response, associated with \textit{shear-thinning} behavior. Taking the bilayer thickness $L_b=4.8$~nm as a characteristic length, the upper limit of validity of the linear response regime (50 bars) can be turned into a surface tension $\phi_u L_b$ of magnitude 25~mN.m$^{-1}$, typical of the oil-water surface tension (35~mN.m$^{-1}$). It corresponds to a typical drift velocity of $10^{-3}$~nm.ps$^{-1}$=1~m.s$^{-1}$. Fig~\ref{fig:Fig7_linear_regime_pull_fluid} represents the average drift velocity $V_u$,as a function of the applied force $F_u$, or equivalently stress $\phi_u=F_u/A$ in the fluid state.

The determination of $V_u$ using CPF and eq.~(\ref{eq:stationaryCouetteFlow}) leads to a value for $b+\eta/L_w$, following eq.~(\ref{eq:ImpulsionFrictionCoefficient}) equal to $\float{2.75\pm 0.08}{6}$~Pa.s.m$^{-1}$. This value was further confirmed by using a larger sample of 1024~lipids with the same hydration of 10 water beads (40 water molecules) per lipid. 
%

%%%
%%%  7.e-4 / 4.e-9 = 1.75e5
%%%

%Note $\eta=\float{7.}{-4}$~Pa.s, $L_w = 3.33$~nm.

%%%%%%%%%%%%%%%%%%%%%%%%%%%
%%% Normalized average displacements fluid
%%%%%%%%%%%%%%%%%%%%%%%%%%%
%%%%%%%%%%%%%%%%%%%%%%%%%%%

%%%%%%%%%%%%%%%%%%%%%%%%%%%
%%% Normalized displacements
%%%%%%%%%%%%%%%%%%%%%%%%%%%
%%%%%%%%%%%%%%%%%%%%%%%%%%%

%%%
\subsection{Fluid phase force kick relaxation (FKR) Couette simulations}

%%%%%%%%%%%%%%%%%%%%%%%%%%%
%\begin{figure}[ht]
%\centering
%\includegraphics[width=0.46\textwidth,angle=0]{Fig8_bootstrap_displacement_fkr_fluid.eps}
%\caption{Ten different bootstrap realizations of the displacement $\mean{\XX_u}(t)$ for an initial kick impulsion $V_0=0.09~\mathrm{nm.ps}^{-1}$ (upper leaflet).}
%\label{fig:Fig8_bootstrap_displacement_fkr_fluid}
%\end{figure}
%%%%%%%%%%%%%%%%%%%%%%%%%%%

%%%%%%%%%%%%%%%%%%%%%%%%%%%
\begin{figure}[ht]
\centering
\includegraphics[width=0.46\textwidth,angle=0]{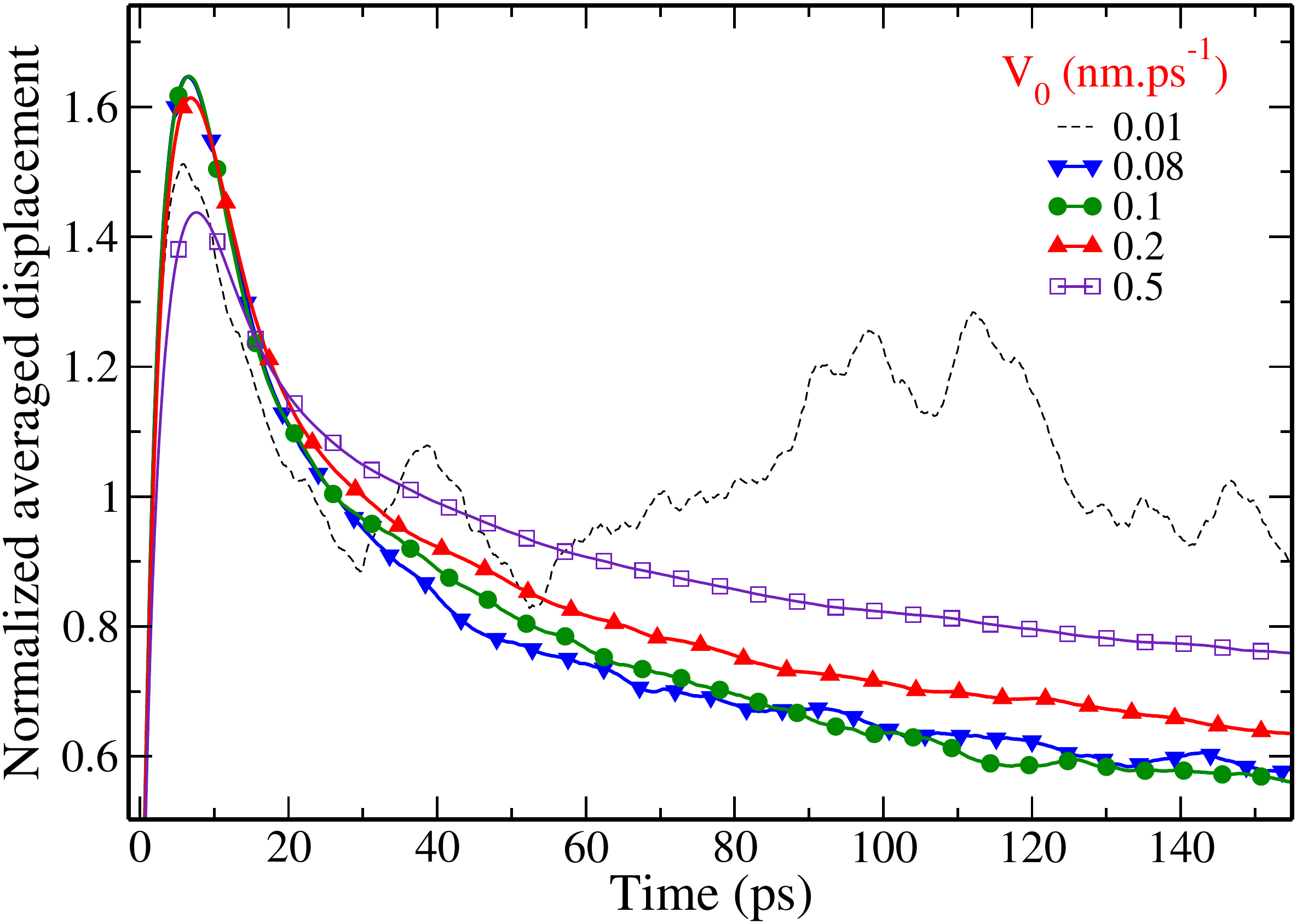}
\caption{Normalized averaged displacements $\mean{\XX_u}(t)/V_0$ for a set of increasing velocities $0.01$ and $ 0.08,\ldots,0.5$ nm.ps$^{-1}$.}
\label{fig:Fig8_normalized_displacement_fkr_fluid}
\end{figure}
%%%%%%%%%%%%%%%%%%%%%%%%%%%

%%%%%%%%%%%%%%%%%%%%%%%%%%%
\begin{figure}[ht]
\centering
\includegraphics[width=0.48\textwidth,angle=0]{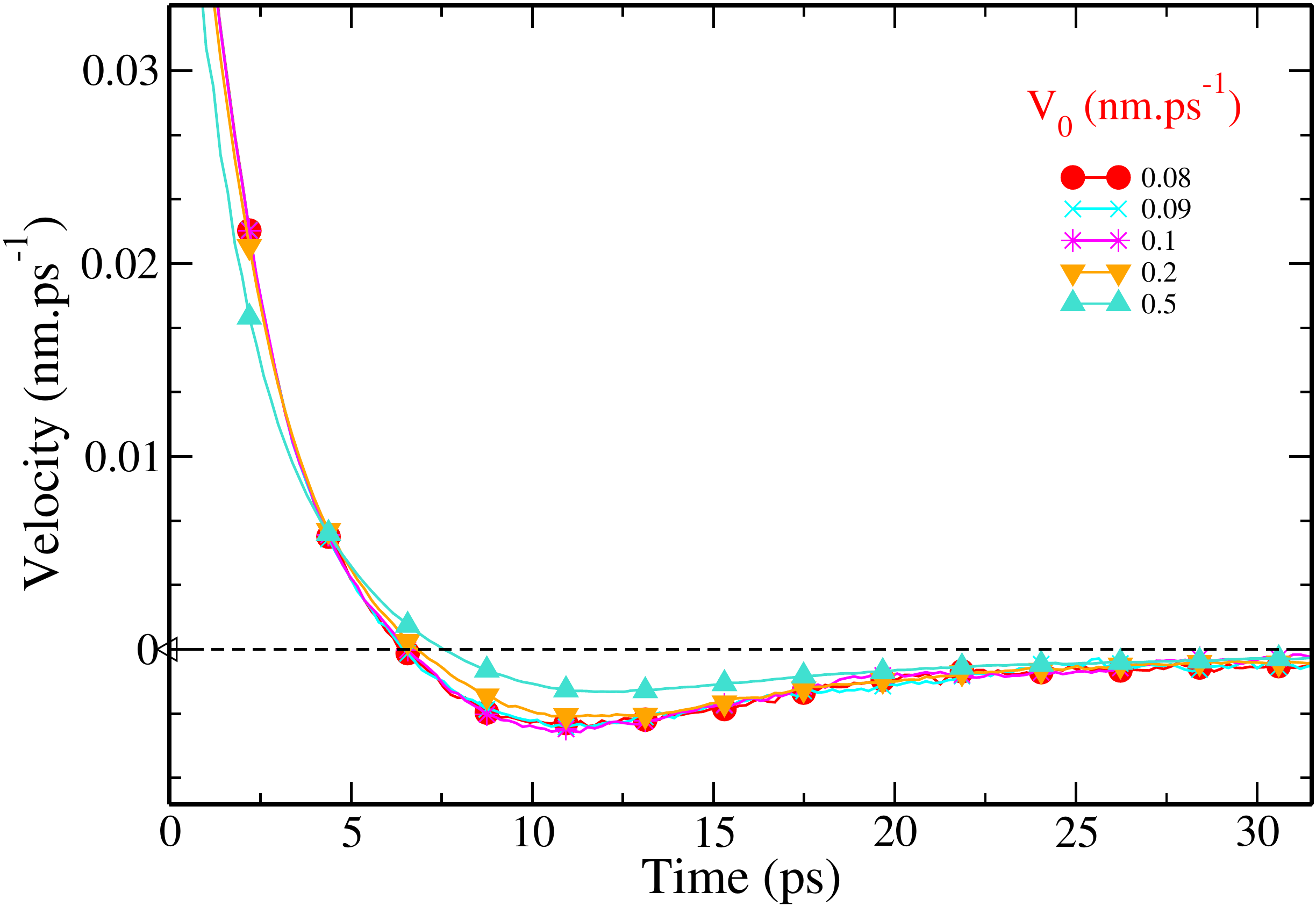}
\caption{Normalized averaged velocities $\mean{\VV_u}(t)=\mean{\mathrm{d}{\XX}_u/\mathrm{d}\,t }(t)/V_0$. The velocity starts at an initial value of~1, decreases fast to~0 (coinciding with the sharp peak in the displacement curve) reaches a negative minimum and finally slowly regresses to 0 from below, coinciding with the slowly decreasing approach of the displacement plateau value.}
\label{fig:Fig9_normalized_velocities_fkr_fluid}
\end{figure}
%%%%%%%%%%%%%%%%%%%%%%%%%%%

%%%%%%%%%%%%%%%%%%%%%%%%%%%
%\begin{figure}[ht]
%\centering
%\includegraphics[width=0.48\textwidth,angle=0]{Fig11_linear_regime_fkr.eps}
%\caption{Displacement $\mean{\Delta \XX_u}$ as a function of initial velocity increment $V_0$ in the Couette force kick relaxation of a fluid bilayer.}
%\label{fig:Fig11_linear_regime_fkr}
%\end{figure}
%%%%%%%%%%%%%%%%%%%%%%%%%%%

Repeated kicks were there applied, starting from 150 to 1000 different configurations. A bootstrap sample of both leaflet displacements is shown in Fig~S1, SI.  The typical averaged displacement curve $\mean{\XX_u(t)}$ increases first linearly, as a natural consequence of the initial force kick that confers a uniform translation velocity to the leaflet (Fig~\ref{fig:Fig8_normalized_displacement_fkr_fluid}). The initial impulsion dissipates fast and vanishes within 5~ps. Surprisingly, the displacement curve starts to decrease, or equivalently the leaflet velocity becomes negative. This peak is followed by a much slower relaxation to an apparent plateau value, also associated with a negative velocity, which extends on a few hundred ps. The apparent plateau value is associated to a relaxation time $t_{\mathrm{relax}}$ such that $\Delta X_u \simeq \mean{\XX_u(t_{\mathrm{relax}})-\XX_u(0)}$, with $t_{\mathrm{relax}}$ of the order of 500~ps. In what follows, for each run $\mathcal{X}^{(\alpha)}(t)$, an estimate of the plateau value was obtained by averaging the displacements over a time interval [500-1000~ps]. 

The striking main feature of the impulsion relaxation curve is the non monotonic behavior of the displacement $X_u(t)$ (Fig~\ref{fig:Fig8_normalized_displacement_fkr_fluid}) and the velocity $V_u(t)$ (Fig~\ref{fig:Fig9_normalized_velocities_fkr_fluid}). It is not possible to account for such a behavior without an elastic contribution to the membrane relaxation. Figs~\ref{fig:Fig8_normalized_displacement_fkr_fluid} and \ref{fig:Fig9_normalized_velocities_fkr_fluid} therefore suggest that the mechanical response of a sheared bilayer is viscoelastic on a time scale  $t_{\mathrm{vel}}\sim t_{\mathrm{relax}}$, with $t_{\mathrm{vel}}$ a bilayer internal viscoelastic relaxation time.  

As in the constant pulling force experiments, it is possible to define a linear response regime, by plotting the displacement normalized with the initial velocity $\mean{\XX_u(t)}/V_0$ as a function of time. A master curve  $\Xi_{(V)}(t)$ is expected to describe this averaged, normalized displacements in the short and intermediate time regime $t\leq t_{\mathrm{relax}}$. The normalized displacement velocity $\dot{\XX}_u/V_0=\VV_u/V_0$ is dimensionless, and can be interpreted as a velocity autocorrelation linked to the momentum scattering efficiency of the mutual interleaflet molecular interactions. 

Fig~\ref{fig:Fig9_normalized_velocities_fkr_fluid} describes the normalized velocity relaxations $\VV_u(t)/V_0$ for a set of increasing $V_0$, and shows a deviation of the relaxation from the master curve $\Xi_{(V)}(t)$ at $V_0$ larger than 0.1-0.2~nm.ps$^{-1}$. Correspondingly, the effective normalized translation shift (plateau) $\Delta X_u(t_{\mathrm{relax}})/V_0$ starts to increase, pointing again to a shear-thinning behavior. The empirical upper bound $V_{l.r.}$ of the linear response regime is therefore found to be of the same magnitude as the maximal velocity $V_{\mathrm{max}}$ deduced from eq.~(\ref{eq:MaximalVelocity}).

While the convergence to a finite plateau value is a reasonable expectation for the averaged displacement curve, simulated trajectories are subject to the thermal motion of the leaflet center of mass, which is expected to be asymptotically dominant at large times. Given a sample size $N_s$, the thermal motion of the sample averaged displacement curve is set to scale as $(D_{\mathrm{COM,u}}/N_s)^{1/2} t^{1/2}$. The determination of $\Delta X_u$ from MD sampling is therefore empirical to a certain extent, as any finite sample average eventually departs from the plateau value. The sample size must be large enough to keep the combination $(D_{\mathrm{COM,u}}/N_s)^{1/2} t_{\mathrm{relax}}^{1/2}$ smaller than $\Delta X_u$. Equation~(\ref{eq:heuristicEinsteinRelation}) provides a theoretical estimate of the accuracy of $\mean{\Delta \XX_u}$. 
The bootstrap estimate of the variance of $\Delta X_u$ (eq.~\ref{eq:bootstrapVariance}) is another independent path to estimate the sample dependence of $\Delta X_u$. 

%%%
\subsection{Fluid phase Poiseuille flow geometry}

Constant pulling rate experiments can be performed in the Poiseuille geometry, when both leaflets are pulled in one direction and the solvent homogeneously pulled in the reverse direction. Assuming that the solvent does not slip at the lipid-solvent interface, the average relative drift velocity obeys relation~(\ref{eq:PoiseuilleStationary}). We justify this assumption from~\cite{2014_Falk_Loison} which found no significant sliding velocity at the lipid water interface on a qualitatively similar system.  Using $L_w=3.5$~nm in the fluid state ($T=340$~K), one finds a value of the coarse grained Martini water viscosity $\eta=\float{8}{-4}$~Pa.s (eq.~\ref{eq:PoiseuilleStationary}). Repeating the simulation  with a larger number of water beads (10240 solvent beads for 512 lipids, $L_w=7.2$~nm), the resulting water viscosity changes to $\eta=\float{7}{-4}$~Pa.s. Independent simulations using reverse non-equilibrium molecular dynamics~\cite{1999_MullerPlathe} (with Lammps, using an equivalent fluid of truncated Lennard-Jones particles at the same temperature) confirms that the solvent viscosity lies close to $\eta=\float{7}{-4}$~Pa.s. The slightly larger value obtained in the presence of a thin water layer is likely to be due to water interfacial effects, the dissipation properties in the interfacial water region being likely to slightly differ from the bulk. The Poiseuille flow simulation design described above can therefore be considered as a viable route to estimate the viscosity of a solvent, provided interfacial effects are small. It is worth noting that the Martini water viscosity  lies quite close to the experimental value, a feature hardly expected from a coarse grained unrealistic water model.

%%%%
\subsection{Gel phase CPF and FKR  simulations}

\begin{figure}[ht]
\centering
\includegraphics[width=0.5\textwidth,angle=0]{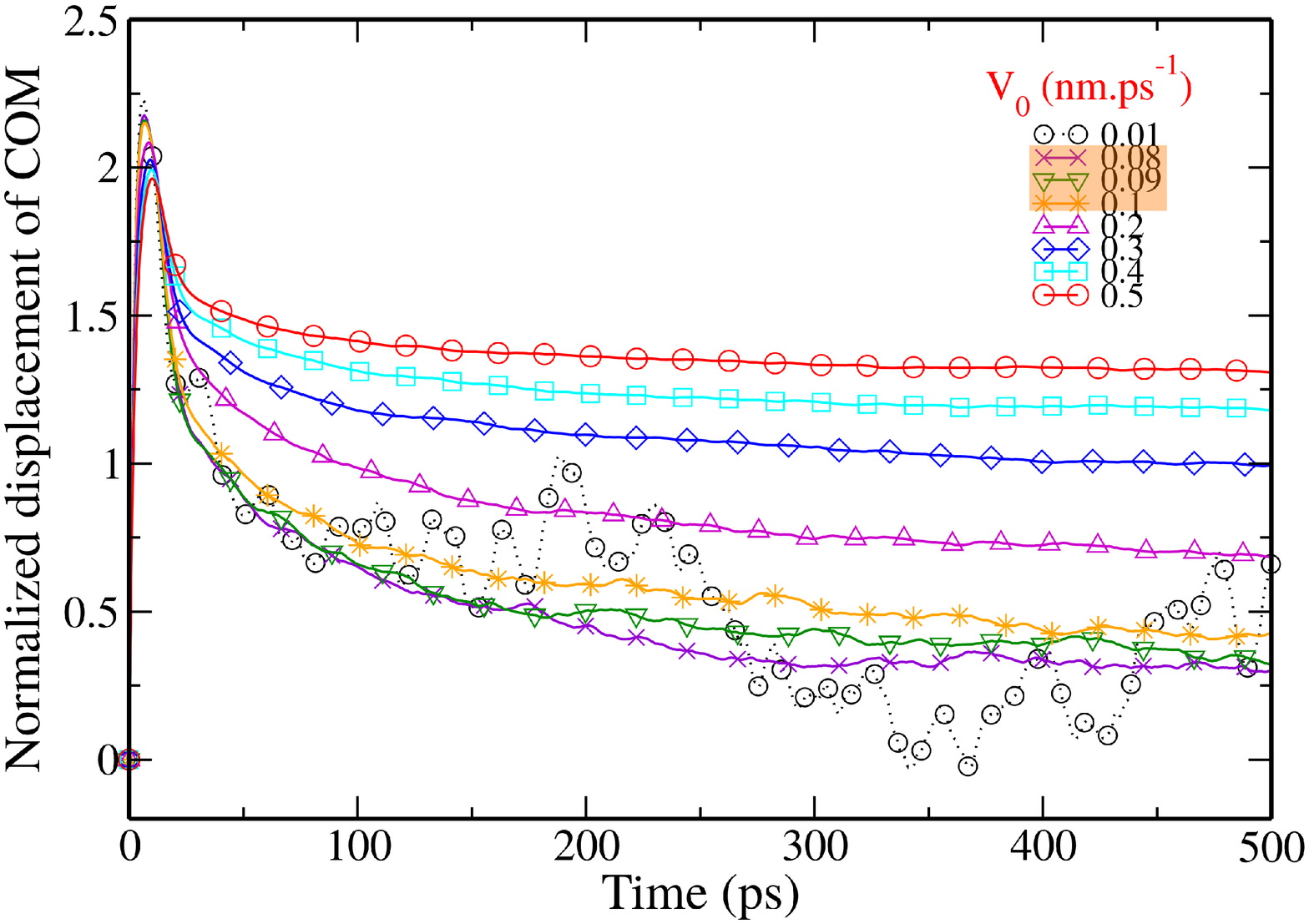}
\caption{Normalized averaged displacements $\mean{\XX_u}(t)/V_0$ in the gel state for a set of increasing impulsions $0.01$ and $ 0.03,\ldots,0.1$ nm.ps$^{-1}$. The plateau value is clearly increasing with the initial applied velocity, and the normalized displacements do not appear to collapse onto a master curve, pointing to an absence  of linear response.}
\label{fig:Fig10_normalized_displacement_fkr_gel}
\end{figure}

\begin{figure}[ht]
\centering
\includegraphics[width=0.48\textwidth,angle=0]{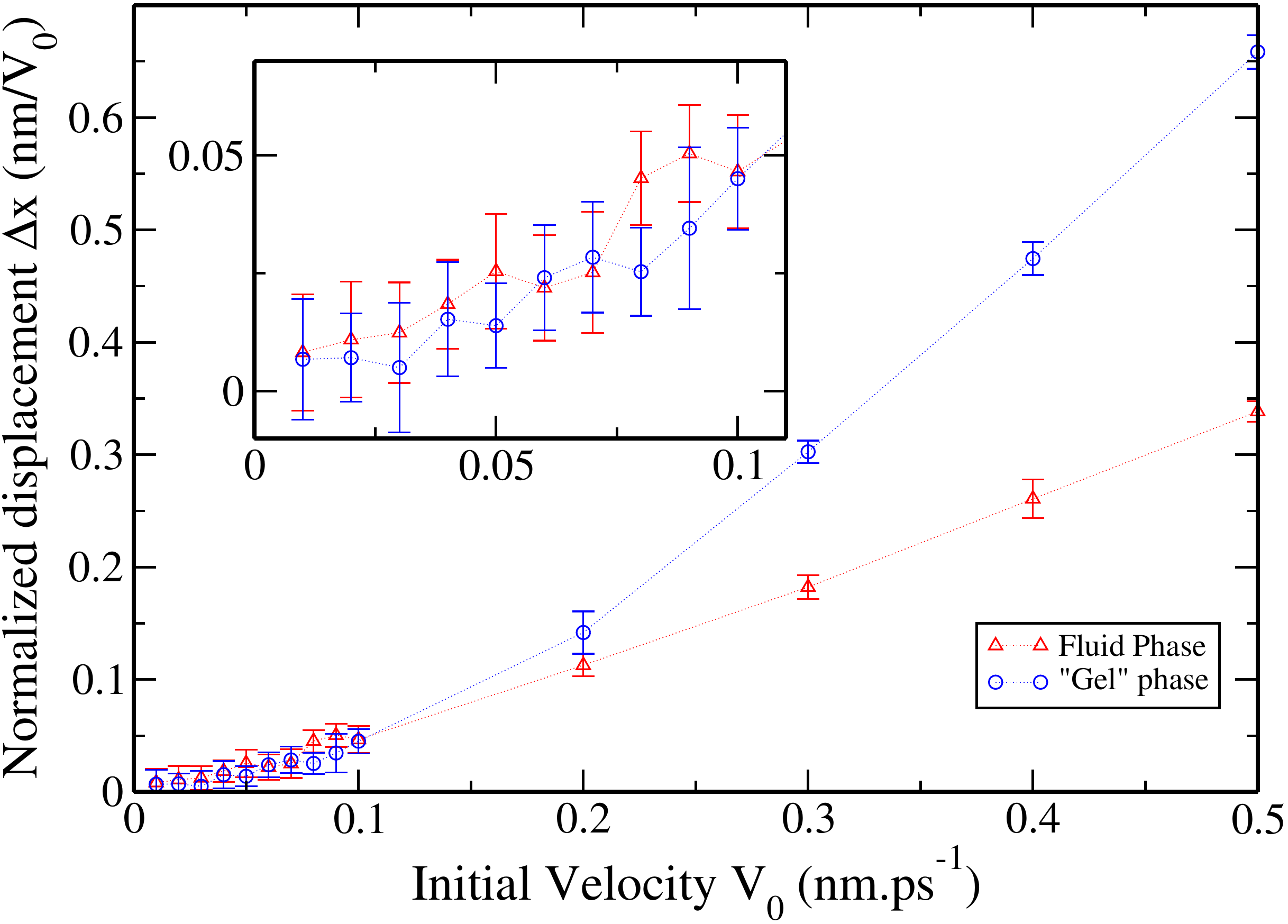}
\caption{Normalized average displacements $\Delta X_u/V_0$ for a set of increasing impulsions and sample size 150, with confidence intervals.}
\label{fig:Fig11_normalized_applied_displacements}
\end{figure}

An ordered phase of the lipid bilayer was obtained at low temperature $T=280$~K. A number of bootstrap realizations of the displacement $\XX_u(t)$ corresponding to an initial velocity step of $V_0=0.9$~nm.ps$^{-1}$ is shown in Fig~S3, SI. Normalized averaged displacements curves $\mean{\XX_u}(t)/V_0$ are represented in Fig~\ref{fig:Fig10_normalized_displacement_fkr_gel}, for increasing initial velocities ranging from 0.01 to 0.5~nm.ps$^{-1}$. The normalized displacements do not superimpose well, even in the low velocity regime, and a master curve $\Xi_{(V)}(t)$ may not exist at low temperatures. This is especially visible in Fig~\ref{fig:Fig11_normalized_applied_displacements}, where the displacements $\Delta X_u$ are plotted as a function of the initial velocity $V_0$. Unlike the fluid phase, the gel phase curve does not display any established linear regime. 

\begin{figure}[ht]
\centering
\includegraphics[width=0.46\textwidth,angle=0]{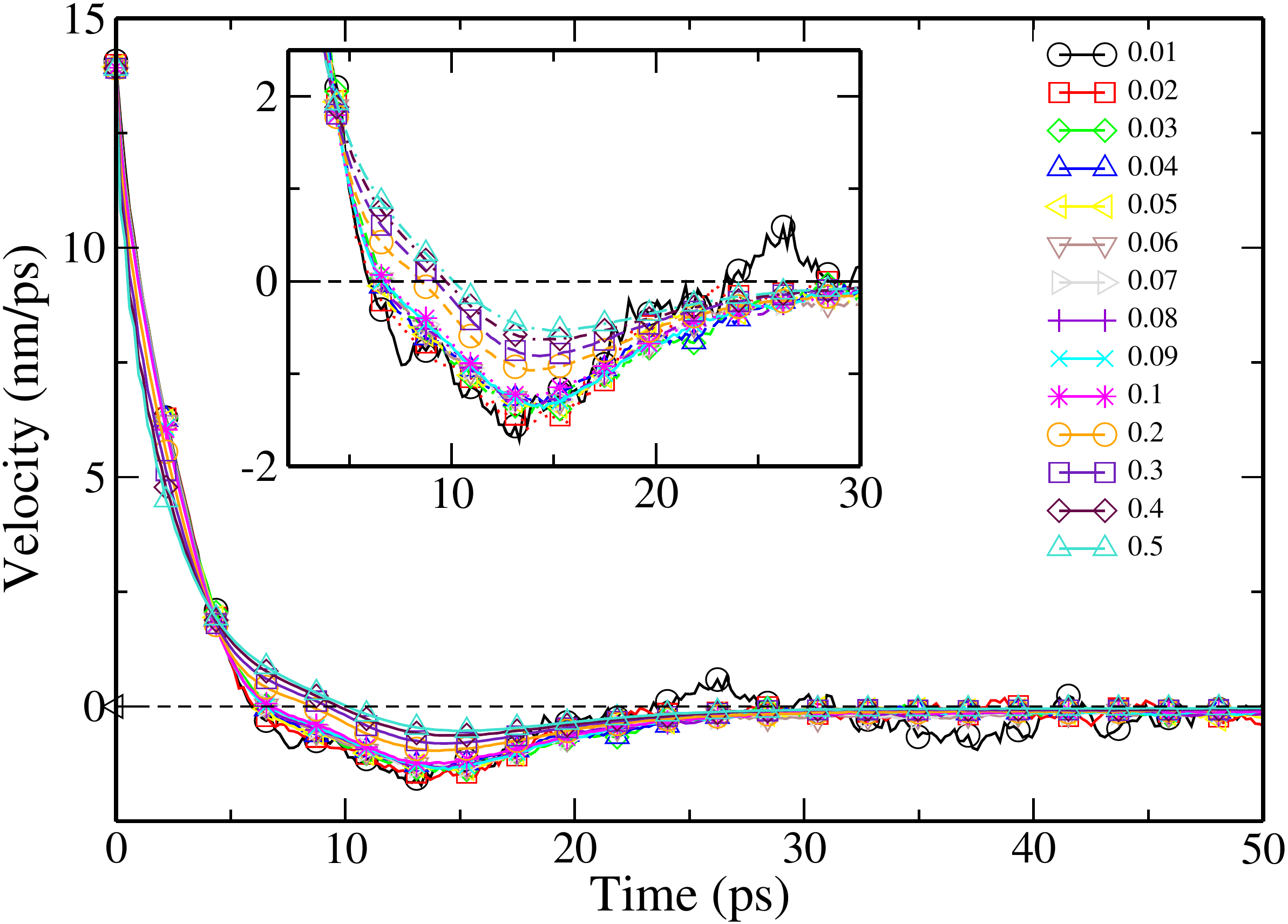}
\caption{Normalized averaged velocity $\mean{\mathrm{d}{\XX}_h/\mathrm{d}\,t }(t)/V_0$. Starting from~1, the normalized velocity crosses 0 fast to reach a minimum, and then relaxes slowly to 0. Curves do not collapse to a master curve. The shape of the normalized relaxation curve is quite different from the fluid case state.}
\label{fig:Fig12_normalized_velocity_fkr_gel}
\end{figure}

Normalized velocities in the low temperature phase are shown in Fig~\ref{fig:Fig12_normalized_velocity_fkr_gel} and Fig~S3, SI
\footnote{Note that due the smaller area per lipid in the gel state, the conversion between applied force $F$ and stress $\phi_u$ is different and now reads $\phi_u = 13400 F$ ($F$ in kJ.mol$^{-1}$.nm$^{-1}$, $\phi_u$ in Pa).}. It is distinctly different from the equivalent fluid counterpart~Fig~\ref{fig:Fig9_normalized_velocities_fkr_fluid}. Correspondingly, the initial displacement peak $\XX_u(t)-\XX_u(0)$ (inset of Fig~S3, SI) is smoother than in the fluid situation. The leaflet velocity change of sign during the relaxation stage is seen both at high and low temperatures. 

\begin{figure}[ht]
\centering
\includegraphics[width=0.5\textwidth,angle=0]{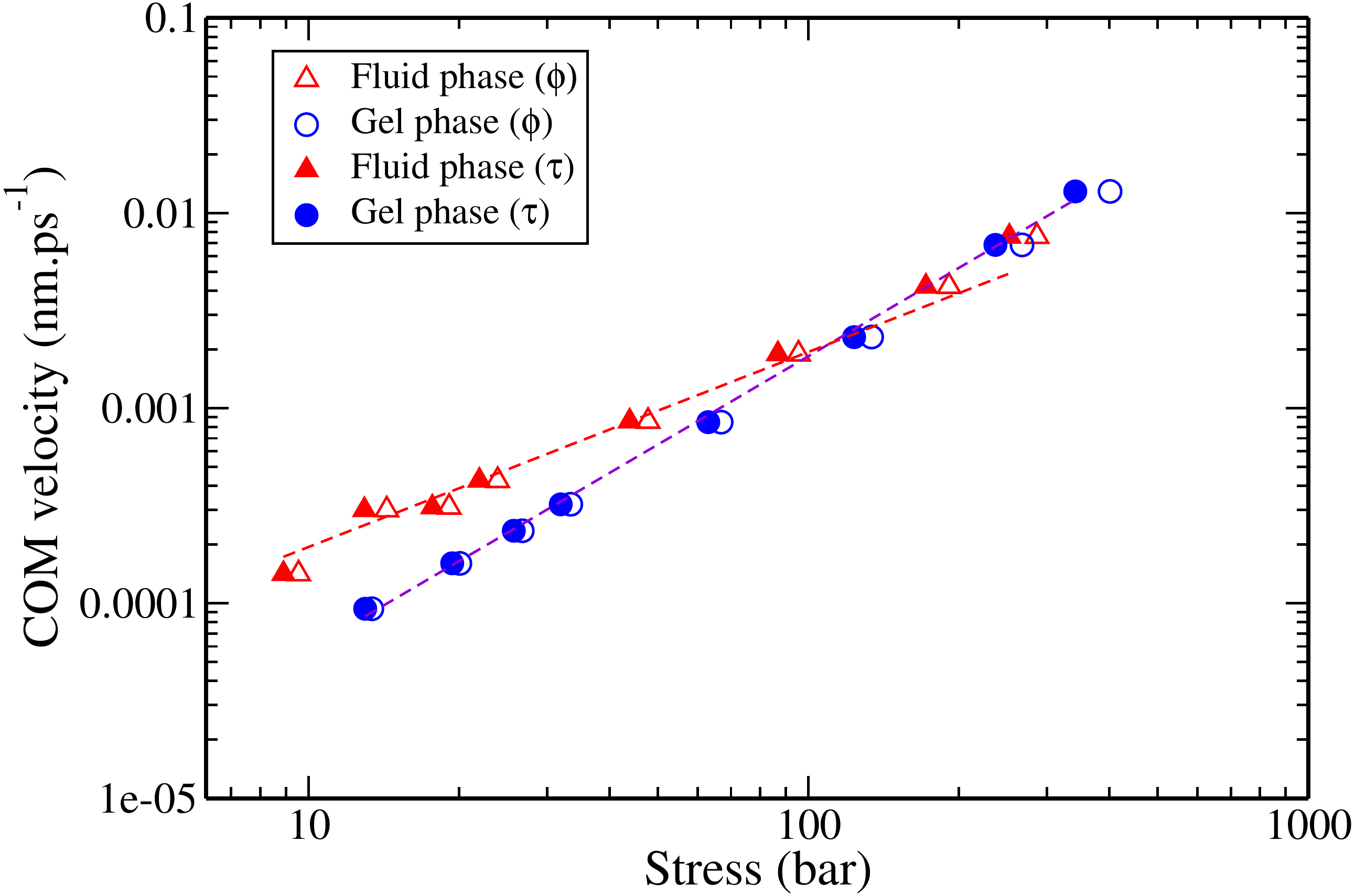}
\caption{Velocity $V_u$- applied shear stress $\phi_u$ and inner stress $\tau$ characteristics in the gel and fluid states, using logarithmic representation. Dashed lines: linear behavior $V_u\sim \tau$ in the fluid state (triangles) and power law behavior in the gel state (circles). The cross-over from linear to shear-thinning regime is visible in the fluid state, while the gel regime seems consistent with a power-low relation of exponent $V_u\sim \tau^{1.50}$ or $\phi_u^{1.44}$.}
\label{fig:Fig13_gel_and_fluid_vs_stress_log}
\end{figure}

In order to extract the true velocity-stress characteristics of the bilayer, we substracted the contribution of the sheared solvent from the applied force. Eq.~\ref{eq:stationaryCouetteFlow} then becomes
\begin{equation}
\tau = \phi_u -\frac{2\eta}{L_w}V_u
\label{eq:innerStress}
\end{equation}
The above relation is valid for an arbitrary stress-velocity relationship, provided the solvent response remains linear in $V_u$. The CPF results in the gel phase are summarized in Fig~\ref{fig:Fig13_gel_and_fluid_vs_stress_log} and Fig~S4, SI. The average drift velocity was plot as a function of the pull force $\phi_u$ and as a function of the inner stress $\tau$. Unlike the fluid phase, the gel phase does not display any linear regime. The log-scale representation of the velocity-stress characteristics seems to indicate a power-low behavior over almost two decades, with apparent exponent $\mean{\VV}\sim\tau^{1.50}$.

%%%%%%%%%%%%%%%%%%%%%%%%%%%
%%% Normalized displacements in the gel phase
%%%%%%%%%%%%%%%%%%%%%%%%%%%
%%%%%%%%%%%%%%%%%%%%%%%%%%%
%%% Normalized velocity in the gel phase
%%%%%%%%%%%%%%%%%%%%%%%%%%%
%%%%%%%%%%%%%%%%%%%%%%%%%%%

%%%%%%%%%%%%%%%%%%%%%%%%%%%%
%%% Normalized average velocity in the gel phase
%%%%%%%%%%%%%%%%%%%%%%%%%%%
%%%%%%%%%%%%%%%%%%%%%%%%%%%

%%%%%%%%%%%%%%%%%%%%%%%%%%%%%%%%%%%%%%%%%%%%%%%%%%%%%%%%%%
%%% Velocity stress characteristics 
%%%%%%%%%%%%%%%%%%%%%%%%%%%%%%%%%%%%%%%%%%%%%%%%%%%%%%%%%%
%\begin{figure}[ht]
%\centering
%\includegraphics[width=0.48\textwidth,angle=0]{Fig16_gel_and_fluid_vs_stress.eps}
%\caption{Constant pull force (CPF) velocity $V_u$- applied shear stress $\phi_u$ and inner stress $\tau$ characteristics in the gel and fluid states (eq.~\protect\ref{eq:innerStress}). If the fluid state display a linear characteristics up to $\tau$=100 bars (linear regime) the characteristics seems nowhere linear in the gel state.}
%\label{fig:Fig16_gel_and_fluid_vs_stress}
%\end{figure}

%%%%%%%%%%%%%%%%%%%%%%%%%%%%%%%%%%%%%%%%%%%%%%%%%%%%%%%%%%
%%% Velocity Stress characteristics in log representation
%%%%%%%%%%%%%%%%%%%%%%%%%%%%%%%%%%%%%%%%%%%%%%%%%%%%%%%%%%
%%%%%%%%%%%%%%%%%%%%%
%%%%%%%%%%%%%%%%%%%%%
%%% MISSING REF
%%%%%%%%%%%%%%%%%%%%%

%%%%%%%%%%%%%%%%%%%%%%%%%%%
\begin{figure}[ht]
\centering
\includegraphics[width=0.18\textwidth]{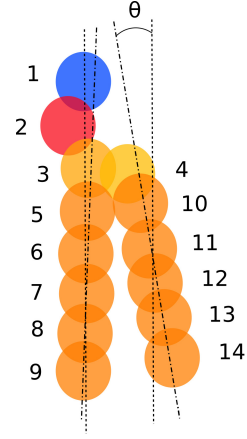}
\caption{Martini CG representation of a DSPC molecule. Bead~1: choline, bead~2: phosphate, beads~3-4: glycerol, beads: 5-9 and 10-14 hydrophobic chains. A vector linking the first and last carbons of each chain is used for defining the average lipid tilt angle $\theta$.}
\label{fig:Fig14_DSPC_Martini} %% Label vide ?
\end{figure}
%%%%%%%%%%%%%%%%%%%%%%%%%%%

%%%%%%%%%%%%%%%%%%%%%%%%%%%%%%%%%%%%%%%%%%%%%%%%%%%%%%%%%%%%
%%%%% Bootstrap coefficient from the impulsion regime
%%%%%%%%%%%%%%%%%%%%%%%%%%%%%%%%%%%%%%%%%%%%%%%%%%%%%%%%%%%%

%%%%%%%%%%%%%%%%%%%%%%%%%%%%%%

\begin{figure}[ht]
\centering
\includegraphics[width=0.5\textwidth,angle=0]{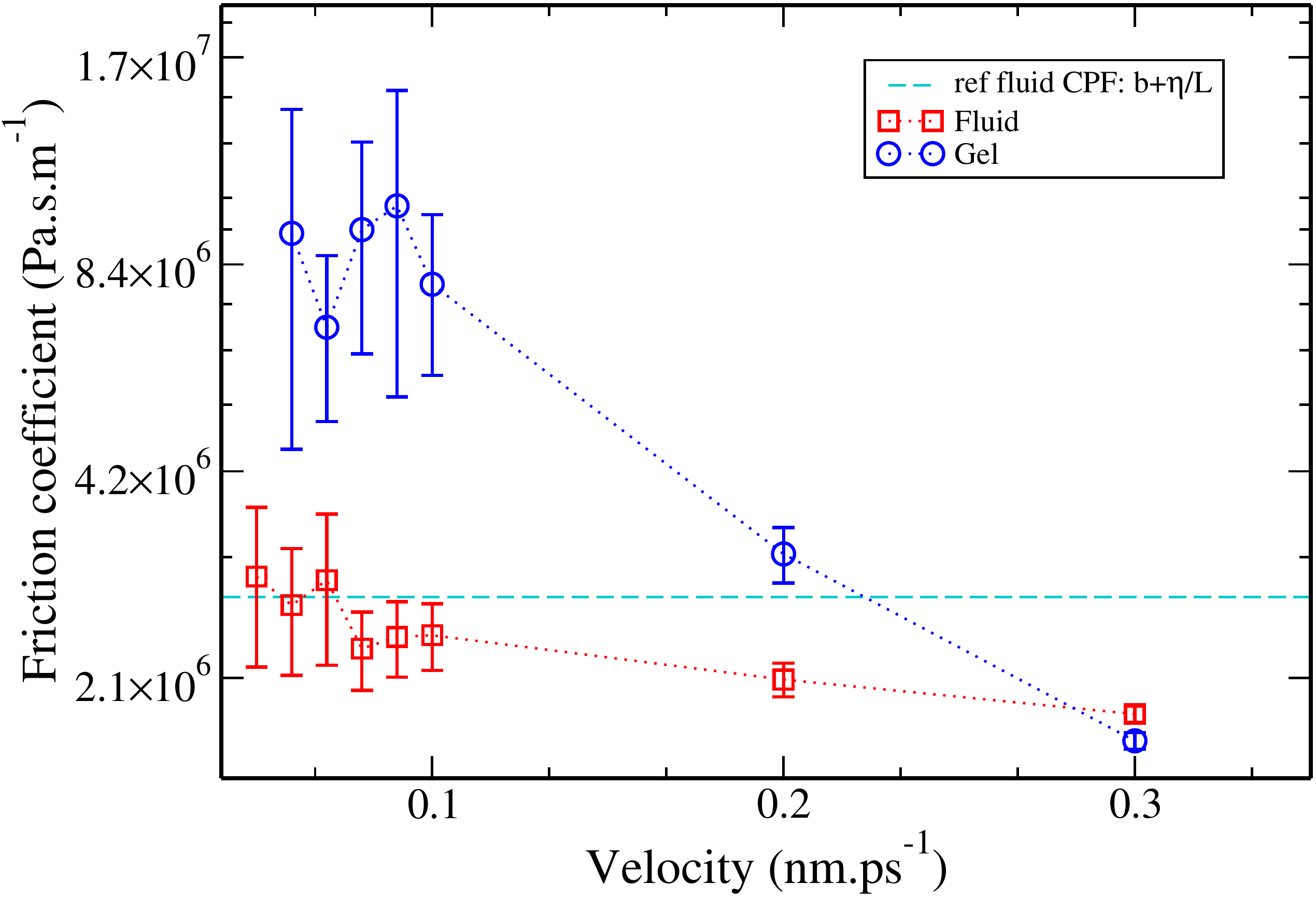}
\caption{Apparent friction coefficient $b+\eta/L_w$ from FKR simulations as a function of the initial induced velocity $V_0$ in the fluid (red squares) and gel (blue circles) states. Each point corresponds to 1000 repeated independent simulations. For each point, a vertical error bar  $2\sigma_b$ is inferred from the bootstrap variance $\sigma_b^2$  of 10 synthetic averaged displacement curves. The resulting confidence interval decreases with velocity in all cases. Confidence intervals are about 10\% relative value for $V_0 = 0.08,0.09$ and 1~nm.ps$^{-1}$, and about 20\% relative value for $V_0=0.05,0.06$ and $0.07$~nm.ps$^{-1}$ in the fluid state.   }
\label{fig:Fig15_fkr_friction}
\end{figure}

%%%
\subsection{Lipid tilt modes}

The non monotonic velocity relaxation curve consecutive to an external force kick at $t=0$ cannot be accounted for by a simple hydrodynamic model. Instead, it suggests that some elastic response is involved in the leaflet translational relaxation. All the numerical evidence suggest that the bilayer remain flat, with negligible out-of-plane bending strain. On the other hand, the simulations are held at constant volume, ruling out standard membrane stretching (or compressibility) contribution. We therefore checked whether lipid tilt modes were activated as a result of the interleaflet friction.  

\begin{figure}[ht]
\centering
\includegraphics[width=0.48\textwidth,angle=0]{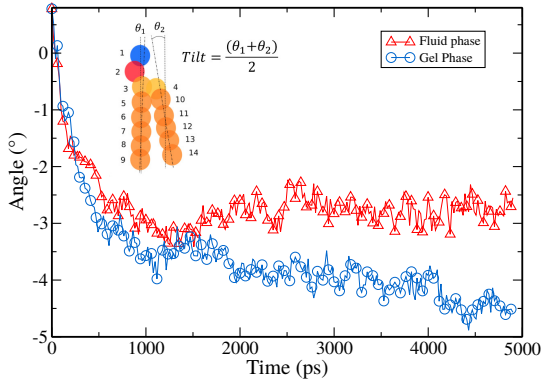}
\caption{Evolution of the average tilt angle $\mean{\theta}$ during a constant pull force  experiment with a force $F=250~\mathrm{kJ.mol}^{-1}.\mathrm{nm}^{-1}$ (stress $\tau=24$~bars).}
\label{fig:Fig16_tilt_cpf}
\end{figure}

We estimated the average lipid tilt angle, defined as  a \textit{polarization} vector linking the first to the least bead in the hydrocarbon chain~(\textit{cf} Fig~\ref{fig:Fig14_DSPC_Martini}). Fig~S5,SI shows, on an enlarged scale, that the average bilayer tilt angle is less than 0.2$^{\circ}$ at equilibrium. When the bilayer is submitted to a CPF, the angle deviates from its vanishing average, proportionally to the applied force (in the limit of linear response and small angles) as shown in Fig~\ref{fig:Fig16_tilt_cpf}. The tilt angle in the fluid phase reaches a well defined asymptotic stationary value, while in the gel phase, the angle seems to be still evolving on the figure time scale (5~ns). In addition, the tilt angle in the gel phase has a larger magnitude than in the fluid phase. The ratio between the average tilt angle and the applied stress is of the order of $\theta/\phi_u\simeq 2.75/24\simeq 0.125^{\circ}.\mathrm{bar}^{-1}$ or $\float{1.7}{-3}$~bar$^{-1}$ with $\theta$ expressed in radians in the fluid phase.

\begin{figure}[ht]
\centering
\includegraphics[width=0.48\textwidth,angle=0]{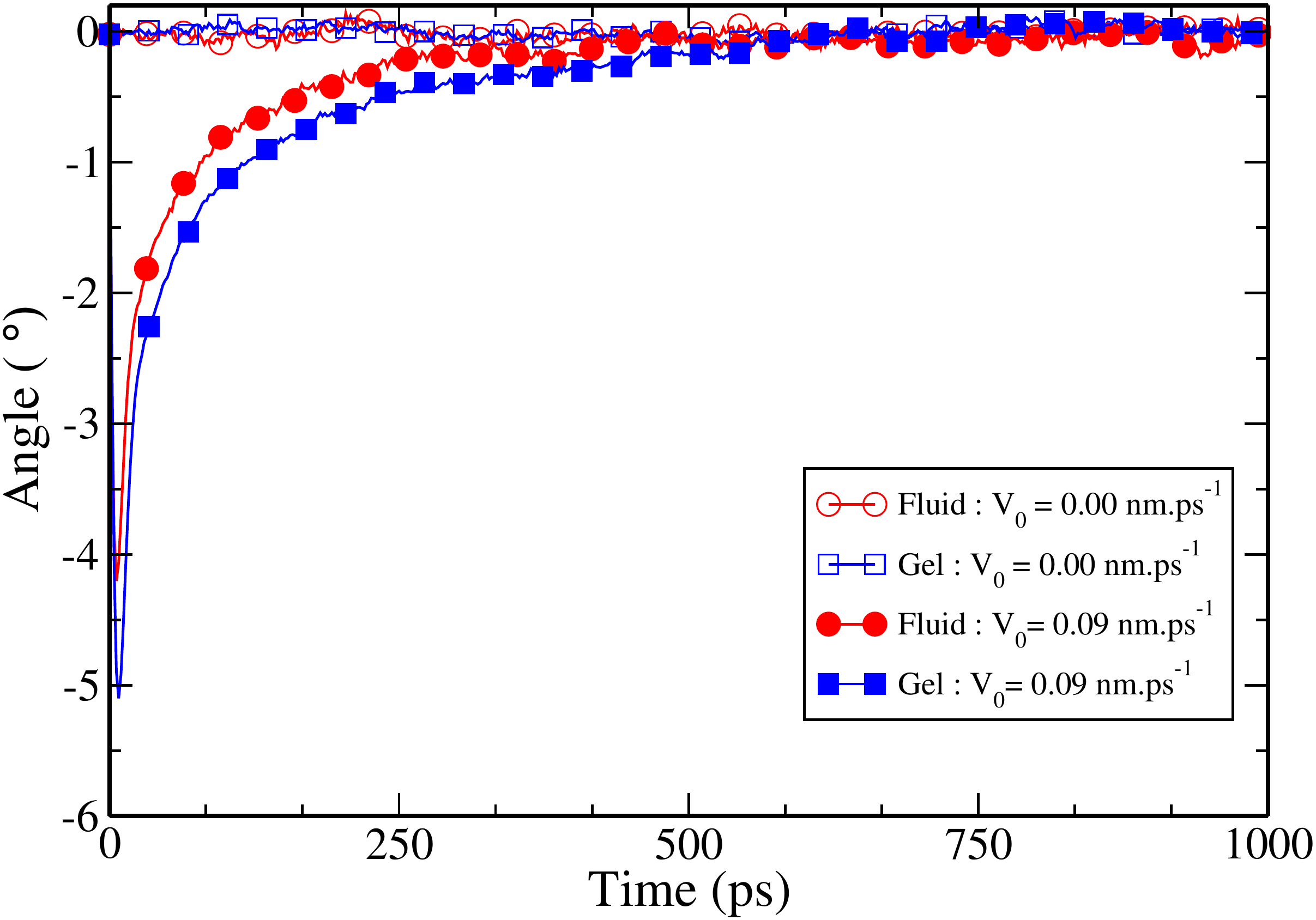}
\caption{Average tilt angle curves $\mean{\theta}$ during a force kick experiment ($V_0=0.09~\mathrm{nm.ps}^{-1}$  in the gel and fluid states respectively).}
\label{fig:Fig17_tilt_fkr}
\end{figure}

Impulsional FKR tilt angle results are shown in Fig~\ref{fig:Fig17_tilt_fkr}, associated to an initial velocity $V_0=0.08$~nm.ps$^{-1}$. For comparison, we also represent equilibrium curves, in the absence of bilayer sollicitation.
The tilt angle in the gel state relaxes slower than in the fluid state. 
%If the fluid phase tilt angle relaxation tends to relax to its vanishing equilibrium value within the time interval considered (600~ps), the gel phase tilt angle does not recede to zero on the same scale.

%Treating each bilayer leaflet as a continuous medium, we measured the average strain between the interfacial and the core bilayer regions. %Independently, we estimated the average lipid tilt angle, defined as a \textit{polarization} vector linking the molecule terminal methyl groups to %the interfacial choline group.  

%%%%%%%%%%%%%%%%%%%%%%%%%%%%%%%%%%%%%%%
%%%% Average Tilt angle CPF
%%%%%%%%%%%%%%%%%%%%%%%%%%%%%%%%%%%%%%%

%%%%%%%%%%%%%%%%%%%%%%%%%%%%%%%%%%%
%%%% Average Tilt angle FKR
%%%%%%%%%%%%%%%%%%%%%%%%%%%%%%%%%%%%

%%%%%%%%%%%%%%%%%%%%%%%%%%%%%%%%%%%%%%%%%%%%%%%%%%%%%%%%%%%%%%
%%% DISCUSSION
%%%%%%%%%%%%%%%%%%%%%%%%%%%%%%%%%%%%%%%%%%%%%%%%%%%%%%%%%%%%%%

\section{Discussion}
\label{sec:Discussion}

The interleaflet sliding kinetics in the fluid phase display an extended linear regime, both in the CPF and the FKR regimes. Deviations from linear behavior emerge as the pulling stress exceeds $\tau_c=50$~bars (Fig~\ref{fig:Fig7_linear_regime_pull_fluid}) or the initial velocity exceeds $0.2$ nm.ps$^{-1}$ (Fig~\ref{fig:Fig9_normalized_velocities_fkr_fluid}). This critical stress $\tau_c$ is of the same magnitude as the cohesion  stresses within the bilayer, of the order of 200~bars~\cite{2009_Ollila_Marrink}. Inversely, small Peclet number considerations make it impractical to use both approaches for too small initial velocities or pull stresses. This lower limit is not intrinsically related to the physical system considered, but a matter of finite simulation box size: increasing the sample size amounts to decreasing the collective center of mass diffusion coefficient and enhances the sensitivity of the method. Unfortunately, unconfined large bilayers systems are subject to strong unfavorable undulation fluctuations, and do not constitute a viable option.

The CPF linear regime yields a consistent estimate for $b+\eta/L_w$, provided one neglects the sliding velocity of the solvent. Substracting off the viscous contribution, our estimate for $b$ is $\float{2.54\pm 0.10}{6}~\mathrm{Pa.s.m}^{-1}$, with $\eta/L_w= \float{2.0\pm 0.12}{5}~\mathrm{Pa.s.m}^{-1}$ obtained from our stationary Poiseuille flow pulling simulations. Different other approaches for the Martini water viscosity ($\float{7}{-4}$~Pa.s.m$^{-1}$ in~\cite{2007_denOtter_Shkulipa}, or using reverse non-equilibrium molecular dynamics with Lammps~\cite{1995_Plimpton}) provides fully consistent estimates. This value compares well with the $A_{55}$ model of den~Otter and Shkulipa obtained using a completely different scheme (RNEMD shear of the surrounding solvent), for which the quoted value for $b$ is between $\float{2.7}{6}$ and $\float{2.8}{6}$~Pa.s.m$^{-1}$. The $A_{55}$ is a similar lipid with 5~beads in each chain (as our DSPC) parameterized using the values of the Martini model, and which was simulated at 323~K. The agreement between both models is very good, given the difference between the approaches and also the 13~K temperature gap. 

Falk et al.~\cite{2014_Falk_Loison} simulated a different coarse-grained model (SDK~\cite{2010_Shinoda_Klein}, see also~\cite{2018_Seo_Shinoda}) obtained a $b$ value of $\float{1.4}{6}$~Pa.s.m$^{-1}$. The difference may be attributed to a difference of parameterization between the SDK and Martini model. This difference is significant enough to change qualitatively the nature of the gel phase. In the SDK model, the low temperature state is a $L_{\beta'}$ tilted chain phase. It results that the SDK solid phase displays anisotropic friction properties, with the direction parallel to the tilt direction displaying a $b$ coefficient close to the fluid case ($\float{1.3}{6}$~Pa.s.m$^{-1}$) and a yield force in the direction perpendicular to the tilt. In our case also, the apparent $b$ value is similar in the gel and fluid case~(Fig~\ref{fig:Fig15_fkr_friction}). However, due to the absence of linear regime, we cannot provide anything but a qualitative behavior of the coefficient~$b$.

Zgorski et al.~\cite{2019_Zgorski_Lyman} performed RNEMD simulations to shear the solvent and the bilayer and obtain $b$, a similar approach as Falk et. al. They compared the old and new version of DPPC Martini lipids (4 beads chains) and obtain a value in the range of $\float{4.3}{6}$ to $\float{5.5}{6}$~Pa.s.m$^{-1}$. These value are significantly larger than ours (even though not strictly comparable) and also than den Otter and Shkulipa ($\float{2.4}{6}$~Pa.s.m$^{-1}$ for the 4 beads chain model $A_{44}$). Interestingly, Zgorski et al. have determined $b$ for the atomistic CHARMM~36 model, reaching values of the order of $\float{1.1}{7}$~Pa.s.m$^{-1}$, still an order of magnitude smaller than the experimental estimates of Evans and Yeung or Pfeiffer et al.~\cite{1994_Evans_Yeung, 1993_Pfeiffer_Sackmann}.
More work is therefore needed, both on the experimental and simulation sides, to determine how accurately current atomistic simulations 
reproduce the local interlayer friction phenomenon.

The FKR predictions for $b+\eta/L_w$ are summarized in Fig~\ref{fig:Fig15_fkr_friction} and the only numerical values in the linear regime with reasonable error bars are those with $V_0 = 0.05$ to $0.1$~nm.ps$^{-1}$.
The resulting confidence interval decreases with velocity in all cases. In the fluid phase, the three first velocities ($V_0=0.05,0.06,0.07~\mathrm{nm.ps}^{-1}$) are consistent with the CPF value obtained in the linear regime (dash line), though with significant error bars. The three following points $0.08,0.09$ and $0.1\,\mathrm{nm.ps}^{-1}$ are located slightly below the CPF value. The last point $0.2\,\mathrm{nm.ps}^{-1}$ is clearly below the CPF value, again pointing towards shear-thinning behavior. 
%\textcolor{blue}{The effective values found in the gel state decreases with temperature, confirming the absence of linear regime, and a power law shear-thinning regime with exponent $b+\eta/L_w\sim V_0^{-0.8}$.}

The estimation of $\Delta X_{\mu}$ used in eq.~(\ref{eq:ImpulsionFrictionCoefficient}) was obtained by computing the average stationary value of the relaxation curves featured in Fig~\ref{fig:Fig8_normalized_displacement_fkr_fluid}. The position of the plateau may have been underestimated as the displacements $\XX_{\mu}(t)$ relaxes slowly to their asymptotic limit. Extending the analysis to longer time scales does not improve much the determination of the displacement because the brownian random diffusion increases, and the signal to noise decreases with the elapsed time. It is therefore necessary to both simulate for longer times and to increase in parallel the number of independent trajectories. We therefore conclude that there is a rough agreement between the CPF and FKR methods. Such an agreement is expected based on linear response considerations, which is only seen in the fluid phase. The numerically observed upper limit of validity of the linear response regime $V_{lr} \sim 0.1~\mathrm{nm.ps}^{-1}$ is remarkably similar to the velocity $V_{\mathrm{max}}\sim 0.2~\mathrm{nm.ps}^{-1}$ deduced from the system kinetic energy argument. This does not directly prove that the excess of kinetic energy is responsible for the linear response breakdown, but it indicates that not other limiting process occurs until the $V_{\mathrm{max}}$ limit is reached.

The FKR approach gives insight on the transient mechanical response of the bilayer, and predicts a sign inversion of the leaflet COM velocity following the positive impulsional initial velocity. We interpret this phenomenon as the consequence of a slowly relaxing lipid chain tilt angle, causing a reactive (non dissipative) stress contribution.  Following the initial velocity kick, an elastic stress builds up, and is further  dissipated. 

We note that a different transient regime would occur if the initial force kick was applied non uniformly to the bilayer leaflets, for instance on the lipid headgroups only. Linear response arguments suggests that the macroscopic hydrodynamic coefficient $b+\eta/L_w$ must not depend on the location of the applied pulling force or force kick. However, the transient response is expected to depend on the way forces are exerted. Further work is needed to compare the current procedure to other possibilities, that would more closely mimic a real shear force pulling experiment.  The uniform pulling force used in the current approach corresponds to a uniform body force applied on each leaflet, due to the fact that all Martini beads have an identical mass. 

A transient shear stress response can be inferred from the retarded memory function formalism exposed in the methodology section. This response can be probed by any spectroscopic shear force experiment, using electromagnetic~\cite{2019_Canale_Bocquet}  or piezoelectric vibrations (dissipative quartz-crystal microbalance QCM-D \cite{1996_Rodahl_Kasemo, 2007_Johannsmann}). So far, none of these techniques reaches the frequency domain of the observed viscoelastic regime. The characteristic "Maxwell" relaxation time predicted by the Martini model is about 100-1000~ps (Figs~\ref{fig:Fig8_normalized_displacement_fkr_fluid},\ref{fig:Fig10_normalized_displacement_fkr_gel}). The connection between Martini coarse-grained and atomistic kinetic properties is quite loose. At room temperature, the Martini lipid diffusion coefficients (ca 70~$\mu$m$^2$.s$^{-1}$ in the DSPC fluid phase at 340~K) are predicted to exceed by a factor 10 the actual values (\textit{ca} 15~$\mu$m$^2$.s$^{-1}$ at 60$^{\circ}$C~\cite{1985_Vaz_Hallmann, Marsh_HandbookLipidBilayers2}). On the other hand, the predicted Martini water viscosity (0.7~mPa.s) is in reasonable agreement with the real value (1~mPa.s). These examples show that the difference between the coarse-grained and atomistic time scales may stretch from 1 to 10, depending on the phenomenon considered. Assuming that the actual relaxation dynamics associated with the leaflet viscoelastic response falls between 1 and 10 times the corresponding numerical prediction, one may estimate the real Maxwell relaxation time to be of order 1--10~ns, and a frequency response  possibly in the 100~Mhz--1~GHz range. 

In addition to the intrinsic membrane elastic response, the water gap probed by the sliding leaflets (Fig~\ref{fig:Fig4_Couette_Poiseuille_flow}) is also expected to respond according to a viscoelastic memory pattern. Stokes hydrodynamics predicts that rigid slabs cannot drag the interstitial fluid instantaneously. The stress-velocity response function can be computed analytically for sticking boundary conditions, using for instance Duhamel's principle~\cite{BookCarlslowJaeger_Heat}. However, if there were no elastic contribution, the viscous memory function alone would not lead to a reversal of the COM velocity. 

The transient response is characterized by a sharp initial increase. We attribute it to the fast loading of the bond springs connecting the beads in the interleaflet area. A characteristic time scale can be obtained as the period $t_{\mathrm{fast}} \sim 5\times 2\pi\sqrt{(m/k_{\mathrm{bond}})}$ of a chain of 5 harmonic spring of stiffness $k_{\mathrm{bond}} = 1250.$~kcal.mol$^{-1}$.nm$^{-2}$ and bead mass $m=$ 72 a.m.u (g.mol$^{-1}$ or atomic mass unit), typical from the Martini force field used in this approach. One finds $t_{\mathrm{fast}}\sim 7$~ps, in reasonable agreement with the observed initial peak dynamics in Fig~\ref{fig:Fig8_normalized_displacement_fkr_fluid}. 

Let us now estimate the hydrodynamic damping time $t_{\mathrm{hyd}}$  resulting from balancing inertia with interleaflet friction. One has $t_{\mathrm{hyd}} = \mathcal{M}N_{l}/(2Ab)$. With $\mathcal{M}\simeq 1000~\mathrm{g.mol}^{-1}$, $N_l =256$, $b = \float{2.5}{6}~\mathrm{Pa.s.m}^{-1}$, $A = (13.2~\mathrm{nm})^2$, we obtain $t_{\mathrm{hyd}}=0.5~\mathrm{ps}$. This time scale est extremely short. We note that it is of the order of magnitude of the normalized plateau value $\Xi_{(V)}(\infty)=\Delta X_u/V_0\simeq 0.5~\mathrm{ps}$ in Fig~\ref{fig:Fig8_normalized_displacement_fkr_fluid}. If the displacement curve following the initial force kick was a single exponential dominated by a balance between friction and inertia, one would see a very fast asymptotic approach to the plateau value, on the same time scale as the first peak. Such a fast relaxation would indeed describe the hydrodynamic response of an incompressible rigid slab subject to solvent and interlayer friction. On the other hand, with $c_t\simeq 1000~\mathrm{km.s}^{-1}$ as the celerity of transverse sound waves in the bilayer (a typical magnitude for a fluid sound wave celerity) it would take at least 5~ps for the sudden shear stress wave following the force kick to establish itself across a 5~nm thick membrane. This proves that the ideal incompressible solid relaxation result cannot describe the observed situation. It also provides an alternative estimate of the characteristic time scale of the initial displacement peak position. 

A Poiseuille characteristic time scale $t_{\mathrm{Poiseuille}}$ can be defined as the slowest relaxation time of the Stokes hydrodynamic flow in a flat slab $\rho L_w^2/(\eta\pi^2)$ involving the channel gap $L_w = 3.5~\mathrm{nm}$ and the water kinematic viscosity $\eta/\rho\simeq \float{7}{-7}~\mathrm{m^2.s}^{-1}$. Its value is $t_{\mathrm{Poiseuille}}=1.6~\mathrm{ps}$, and also much shorter than the observed relaxation time. We therefore conclude that the sliding leaflets relaxation time $t_{\mathrm{relax}}$ has a viscoelastic origin, and we denote it $t_{\mathrm{vel}}$. 

We therefore conclude that the relaxation seen on Fig~\ref{fig:Fig8_normalized_displacement_fkr_fluid} results from slow membrane internal relaxation dynamics and is not limited to the interfacial sliding region. Slow lipid tilt modes relaxation, such as depicted in Fig~\ref{fig:Fig16_tilt_cpf}, certainly contribute to the observed slow viscoelastic response of the bilayer FKR. 

\begin{figure}[ht]
\centering
\includegraphics[width=0.42\textwidth,angle=0]{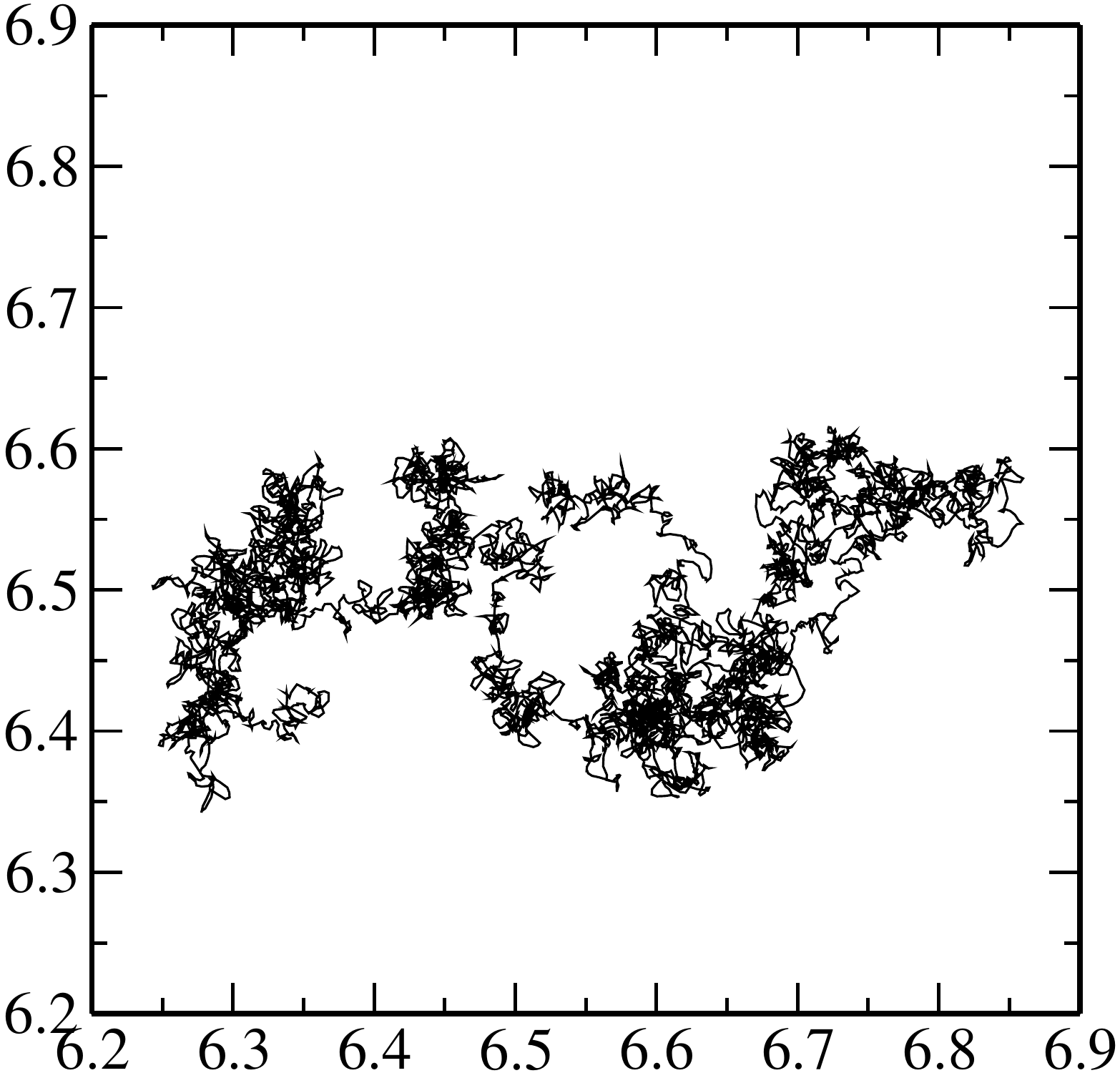}
\caption{Projected Brownian $xy$ trajectory of the upper leaflet center of mass, observed during 50~ns in the fluid phase.}
\label{fig:Fig18_brownian_COM}
\end{figure}

\begin{figure}[ht]
\centering
\includegraphics[width=0.46\textwidth,angle=0]{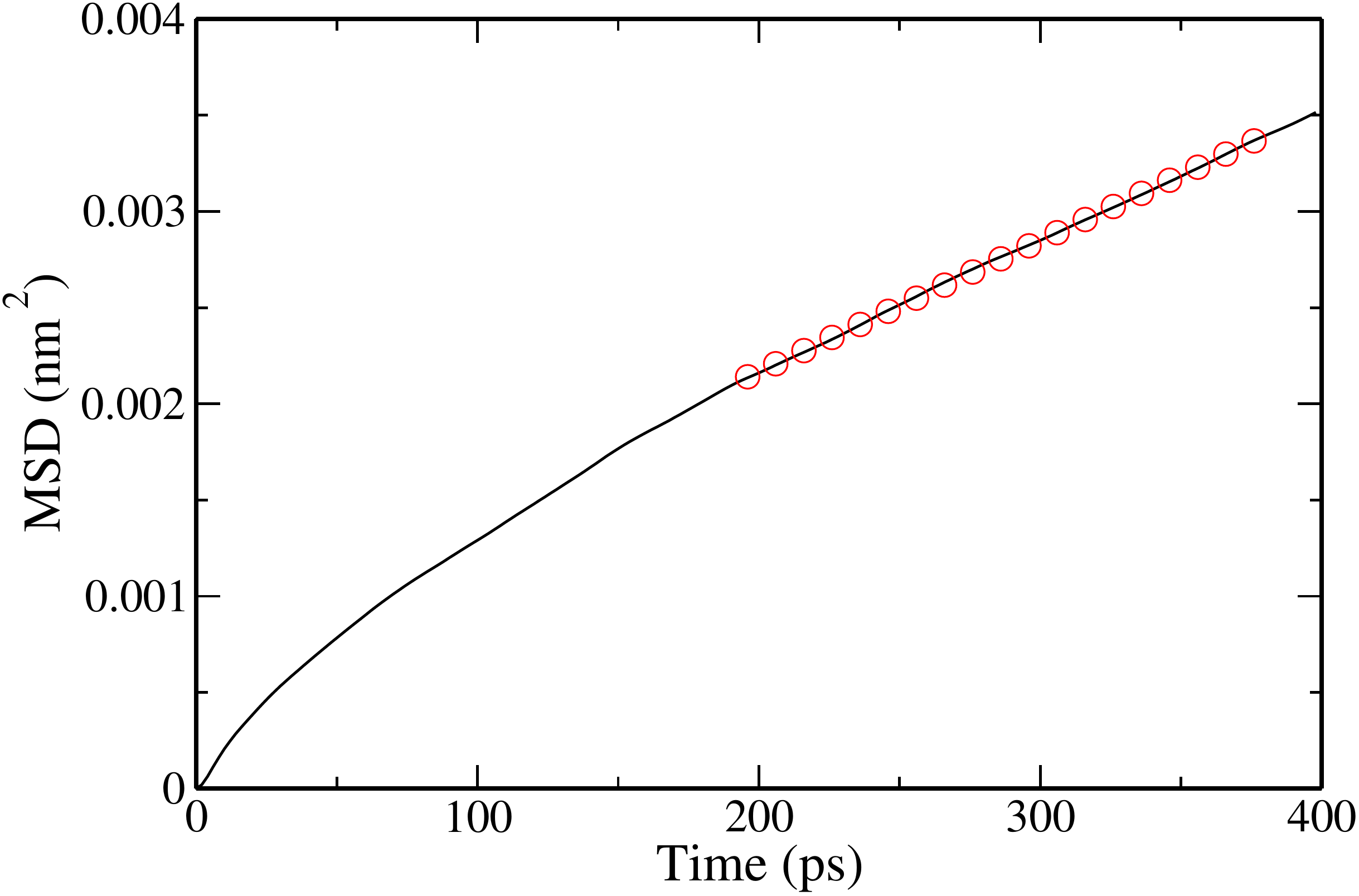}
\caption{Mean square displacement curve of the upper leaflet center of mass at 340~K and a linear adjustment with $2D=\float{6.8}{-5}~\mathrm{nm}^{2}.\mathrm{ps}^{-1}$.}
\label{fig:Fig19_msd_COM}
\end{figure}

It turns out that the condition $\Delta X_u \sim  V_0 t_{\mathrm{relax}}$ is not met. Instead $\Delta X_u/ V_0$ is of the order of $t_{\mathrm{hyd}}\sim 0.5~\mathrm{ps}$  and the long relaxation time $t_{\mathrm{relax}} \sim t_{\mathrm{vel}} \gg \Delta X_u/ V_0$ enhances the effect of brownian fluctuations.
Following eq.~(\ref{eq:heuristicEinsteinRelation}) one expects a COM diffusion coefficient of the order of 10~$\mu$m$^2$.s$^{-1}$. A numerical estimate based on the COM mean squared displacement yields a value $D_{\mathrm{COM,u}} \simeq 3.4~\mu\mathrm{m}^2.\mathrm{s}^{-1}$ (Figs~\ref{fig:Fig18_brownian_COM} and~\ref{fig:Fig19_msd_COM}).

%%%%%%%%%%%%%%%%%%%%%%%%%%%%%%%%%%%%%%%%%%%%%%ù
%%% Sample Brownian trajectory
%%%%%%%%%%%%%%%%%%%%%%%%%%%%%%%%%%%%%%%%%%%%%%

%%%%%%%%%%%%%%%%%%%%%%%%%%%%%%%%%%%%%%%%%%%%%%ù
%%% MSD COM
%%%%%%%%%%%%%%%%%%%%%%%%%%%%%%%%%%%%%%%%%%%%%%

%%%%%%%%%%%%%%%%%%%%%%%%%%%%%%%%%%%%%%%%%%%%%%
%%% No reference

As a consequence, the ballistic to brownian displacement ratio $r_{\mathrm{bal/br}}$ equals  
\begin{eqnarray}
r_{\mathrm{bal/br}} &=& 
    \frac{\Delta \XX_u}{\sqrt{D_{\mathrm{COM,u}} t_{\mathrm{vel}}}}
    \nonumber\\
    &=& \left(\frac{V_0^2 t_{hyd}}{D_{\mathrm{COM,u}}}\right)^{1/2} \left(\frac{t_{hyd}}{t_{vel}}\right)^{1/2} 
\end{eqnarray}
For an initial velocity jump $V_0=0.08~\mathrm{nm.ps}^{-1}$, the ratio $r_{\mathrm{bal/br}}\simeq 1.$ According to the above expression, brownian displacement and ballistic drift are for each single run $\XX^{(\alpha)}(t)$ of the same order of magnitude. With $t_{\mathrm{vel}}\simeq 500$~ps, one finds from Fig~\ref{fig:Fig19_msd_COM} a mean square displacement of the center of mass of the order of 0.004~nm$^{2}$. Assuming a gaussian distribution of the latter and a sample size of $N_s=1000$ independent runs,  the resulting $2\sigma_b$ confidence interval is expected to be $\sim 10\%$, the right order of magnitude for what is seen in ~Fig~\ref{fig:Fig15_fkr_friction}.

We finally note that that the friction coefficient $b=\float{2.54}{6}~\mathrm{Pa.s.m}^{-1}$ can be interpreted as a Newtonian fluid sheared between two infinitely thin parallel planes separated by a 5~nm thick gap, with an equivalent dynamic viscosity $\eta_{\mathrm{equiv}}= 13~\mathrm{mPa.s}$, about 15 times the value of liquid water.

Let us now consider the friction properties of the bilayer in the gel state. The most prominent characteristics is the absence of visible linear response regime. This is particularly clear from Figs~\ref{fig:Fig11_normalized_applied_displacements}, \ref{fig:Fig13_gel_and_fluid_vs_stress_log} and \ref{fig:Fig15_fkr_friction}. The effective $b+\eta/L_w$ coefficients decreases with the external pulling stress (CPF) and the initial force kick (FKR),  a typical shear-thinning behavior. As the solvent viscosity does not change at the transition, the interleaflet friction is responsible for the observed behavior. If it is not possible to affirm for sure that no linear regime exists at lower pulling stresses, such a linear regime clearly lies beyond our current simulation capacities. 

Shear thinning behavior is the hallmark of complex fluids dynamics. In the CPF regime, the effective friction $b$ appears to follow an approximate power-law regime $V_u \sim \tau^{1.5}$, or equivalently $b=\tau/V_u \sim V_u^{-0.33}$, where $\tau$ is the shear-stress. Beyond linear response, one does not expect equivalence between CPF and FKR measurements in the gel phase. 

The tilt relaxation dynamics (Figs~\ref{fig:Fig16_tilt_cpf} and~\ref{fig:Fig17_tilt_fkr}) suggests that the lipid tilt relaxation occurs slowly in the gel phase. A possible explanation would be that irreversible or slowly reversible plastic deformations are involved in the gel sheared bilayer. 
%Fig~\ref{fig:Fig20_plastic} illustrates a possible mechanism. Assuming that straight lipid chains are sufficiently ordered in the gel phase, a stationary tilt angle arises as a consequence of lateral stress. In the absence of lattice defect, Hookian (linear) elasticity is expected from the strained structure. By introducing lattice defects and rotating the structure, it is possible to tilt the lipid chains without, or with smaller elastic stress. Such defects should not be be present in bulk crystalline structures, but become much more likely in a slab of finite thickness. One expects in this case an overall softer elastic response $\tau(\theta)$ of the bilayer which could account for the nonlinear friction response. 
However, we have not yet found a quantitative explanation for the apparent power-law exponent of the velocity-force characteristics. 

%Finally, the induction of a tilt $\theta$ by the applied stress $\tau$ suggests the possibility a normal stress $\tau\theta\sim \float{1.8}{-3}\tau^2 $ in the fluid phase. With $\tau\sim 50$~bars the normal stress is of the order of 5~bars. 

%%%%%%%%%%%%%%%%%%%%%%%%%%%%%
%\begin{figure}[ht]
%\centering
%\includegraphics[width=0.37\textwidth,angle=0]{Fig20_plastic.eps}
%\caption{(a) Putative schematic representation of unstrained lipid chains ordered phase. (b) As a consequence of sliding friction, strain builds up in the gel phase. (c) strain can be decreased by shifting the crystalline network (pairs of edge dislocations) and rotating the resulting "staircase" shape. In (b) and (c) the black full arrow is normal to the local lipid-water interface, the blue full arrow is aligned with the chains, so that the angle between both full arrows is proportional to the local strain. The dashed arrow in (c) is the global bilayer normal.}
%\label{fig:Fig20_plastic}
%\end{figure}
%%%%%%%%%%%%%%%%%%%%%%%%%%%%%%

%%%%%%%%%%%%%%%%%%%%%%%%%%%
%\begin{figure}[ht]
%\centering
%\includegraphics[width=0.46\textwidth]{Figures/Fig-ShearThinningPull.eps}
%\caption{Interleaflet friction coefficients $b$ measured for the upper and lower leaflets for increasing stress values. Shift to the discussion part. /ShearThinningPull}
%\label{fig:ShearThinningPull}
%\end{figure}
%%%%%%%%%%%%%%%%%%%%%%%%%%%

%%%%%%%%%%%%%%%%%%%%%%%%%%%%%%%%%%%%%%%%%%%%%%%%%%%%%%%%%%%%%%
%%% DISCUSSION
%%%%%%%%%%%%%%%%%%%%%%%%%%%%%%%%%%%%%%%%%%%%%%%%%%%%%%%%%%%%%%

\section{Conclusion}
\label{sec:Conculsion}

We investigated two different approaches for studying lipid bilayer friction, which can both be generalized to supported membrane systems. 
A constant pull force method was used to determine the solvent shear viscosity and the bilayer interleaflet friction, with good accuracy. 
A DSPC fluid membrane was found to behave linearly until the shear stress reaches the order of 50-100~bars, and the sliding velocity the order of 1 to 2~m.s$^{-1}$. Meanwhile in the gel state no linear response was observed, but instead a non-linear power law stress velocity characteristics. The magnitude of the friction is similar in both phases. 

A second original approach consisted in monitoring the relaxation of the membrane drift motion following an initial force kick. This method was found to be less accurate, but consistent with the previous one. It reveals that both fluid and gel membranes relax slowly to equilibrium, on a characteristic time scale $t_{vel}$ much larger than the hydrodynamic hydrodynamic damping $t_{\mathrm{hyd}}$. The overall response bears the hallmark of linear viscoelasticity.

The next step will consists in applying the pull and kick force methods to atomistic models of fluid bilayers, and to supported bilayer membranes where strong confinement and interaction between solid surface and bilayer may change significantly the results.  

The authors warmly thanks Tiago E. de Oliveira, Adrien Gola and Olivier Benzerara for help and discussions, and gratefully acknowledge support from the high performance cluster (HPC) Equip\@Meso from the University of Strasbourg, through grant n$^{\circ}$G2018A53.

%%%%%%%%%%%%%%%%%%%%%%%%%%%%%%%%%%%%%%%%%%%%%%%
%%%% BIBLIOGRAPHY
%%%%%%%%%%%%%%%%%%%%%%%%%%%%%%%%%%%%%%%%%%%%%%%

%\section{Bibliography}
%\bibliography{biblio_sliding}

%%%%%%%%%%%%%%%%%%%%%%%%%%%%%%%%%%%%%%%%%%%%%%%
%%%  APPENDIX
%%%%%%%%%%%%%%%%%%%%%%%%%%%%%%%%%%%%%%%%%%%%%%%

%\newpage
\appendix

%%%%%%%%%%%%%%%%%%%%%%%%%%%%%%%%%%%%%%%%%%%%%%%
%%% SIMULATION DETAILS
%%%%%%%%%%%%%%%%%%%%%%%%%%%%%%%%%%%%%%%%%%%%%%%

\section{Simulation details}
\label{app:SimulationDetails}

We used the Martini lipid version \textit{v2.0} and \textit{Gromacs~5.1}. The representation of a DSPC lipid is described in Fig~\ref{fig:Fig14_DSPC_Martini}. It consists in 14~beads located at various levels on an hydrophilicity scale, interacting with Lennard-Jones interactions of radius $r_0=0.47$~nm,  connected with harmonic springs of stiffness $k_0=1250$~kJ.mol$^{-1}$~\cite{2004_Marrink_Mark,2007_Marrink_deVries}.

In all the simulations, the standard Gromacs \textit{md} leap-frog molecular dynamics integrator was used, with a time step of 20~fs. The velocity rescale~\cite{2007_Bussi_Parrinello} was used to keep the energy constant in the simulation. This thermostat is an alternative to  Nose-Hoover and uses a single supplementary stochastic coordinate $Q$ ensuring canonical ensemble ergodicity for the simulated system. Lipid and solvent groups of molecules were separately coupled to two v-rescale thermostats, with a coupling time constant of 1~ps. For constant pressure simulations, we used a semi-isotropic Parinello-Rahman barostat with a time coupling constant of 12~ps and  a compressibility $\float{3}{-4}~\mathrm{bar}^{-1}$ in the $xy$ and $z$ directions. 

Center of mass (COM) fixation (\textit{nstcomm}) deserves a special attention. It is required to fix the system COM to a constant position as soon as the system in translation invariant conjugated with the use of a Nose-Hoover or v-rescale thermostat. In the Couette flow situation, the bilayer and water groups have a separately vanishing linear momentum. In the Poiseuille flow, only the system COM is stationary. One must therefore apply the constraint on the system center of mass (which would otherwise not be perfectly steady due to the approximate treatment of intermolecular forces), and not separately to the subsystems. 

%%%%%%%%%%%%%%%%%%%%%%%%%%%
\begin{figure}[ht]
\centering
\includegraphics[width=0.45\textwidth]{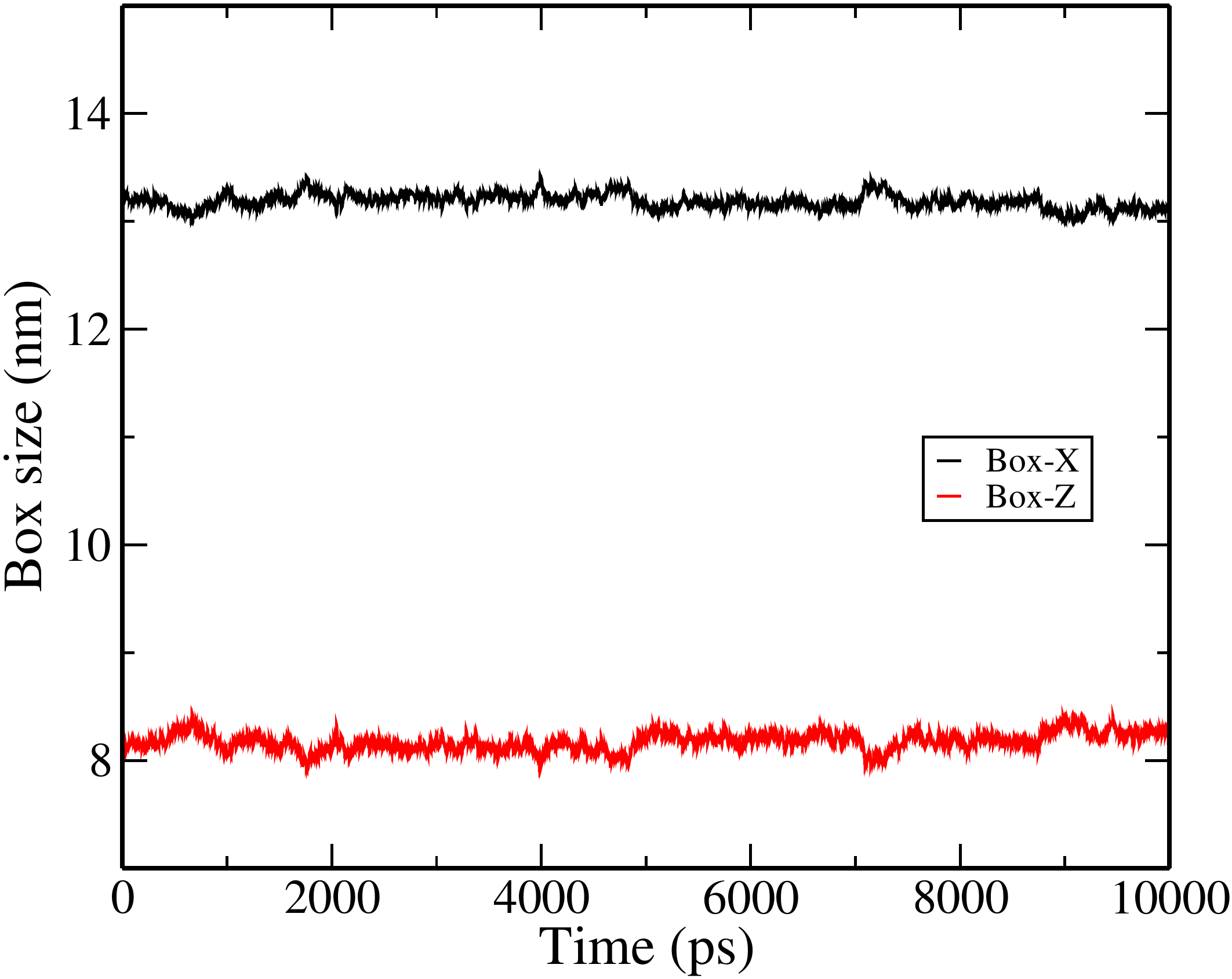}
\caption{Evolution of the horizontal and vertical box sizes during the NPT simulation, used for determining the average box size, in the fluid phase.}
\label{fig:Fig20_box_fluid} 
\end{figure}
%%%%%%%%%%%%%%%%%%%%%%%%%%%
%%% NON REFERENCE
%%%%%%%%%%%%%%%%%%%%%%%%%%

%%%%%%%%%%%%%%%%%%%%%%%%%%
\begin{figure}[ht]
\centering
\includegraphics[width=0.45\textwidth]{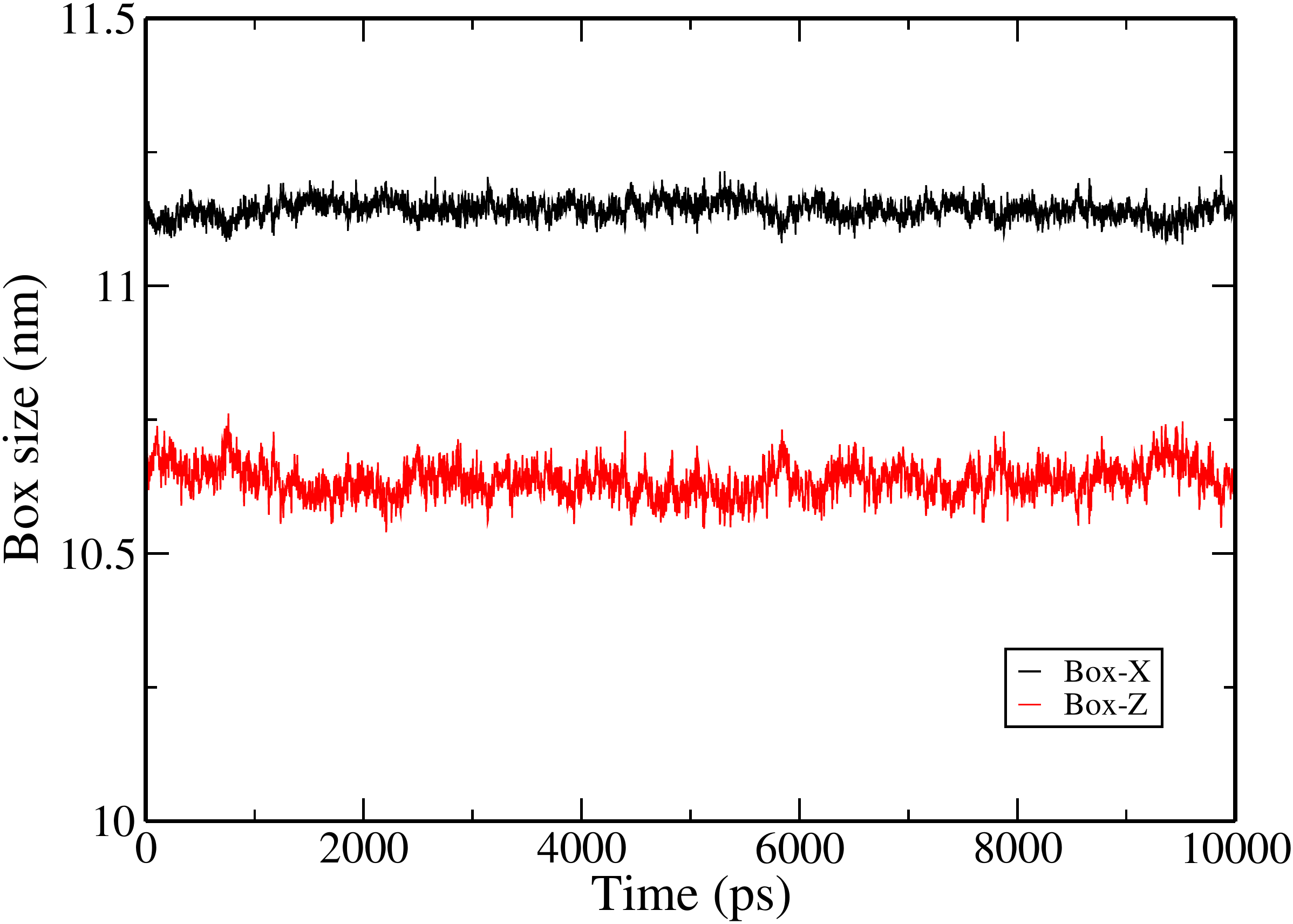}
\caption{Evolution of the horizontal and vertical box sizes during the NPT simulation, used for determining the average box size, in the gel phase.}
\label{fig:Fig21_cold_box} 
\end{figure}
%%%%%%%%%%%%%%%%%%%%%%%%%%%

A NPT run of 40~ns was used to determine the average box size for a system subject to constant pressure conditions (Fig~\ref{fig:Fig20_box_fluid}, \ref{fig:Fig21_cold_box}). A NVT run of 1~$\mu$s was then used to generate 1000 thermalized initial conditions, both in the fluid and the gel phases. 

Constant force pulling was implemented using umbrella sampling control parameters, such as in the following example for pulling in the Couette geometry with a constant force of 250~kJ.mol$^{-1}$.nm$^{-1}$:
\begin{verbatim}
Pull =yes
pull_ngroups = 2
pull_ncoords = 1
pull_group1_name = up
pull_group2_name = down
pull_coord1_type = constant-force
pull-coord1-vec = 1 0 0
pull_coord1_geometry =  direction-periodic  
pull_coord1_groups = 1 2
pull_coord1_dim = Y N N 
pull_coord1_k  = 250
pull_coord1_start  = yes
\end{verbatim}
In the CPF analysis, 50 trajectories of 10~ns were used and combined for each pulling stress condition. 

Force-kick relaxation simulations were realised by changing with a python script the $x$ components of the velocities in the initial configuration file (\textit{gro} file when using Gromacs 5.1) as suggested in eq.~\ref{eq:InitialVelocityShift} and using the new velocities as a starting configuration. The FKR relaxation dynamics is unusual in terms of short characteristic relaxation times, of the order of 1~ps. To perform our statistical analysis, trajectory frames were dumped every 10 time-steps (0.2~ps) and 150~trajectories of 25000 steps (500~ps) were used and combined for each initial velocity condition. Home-made analysis software was used to open and extract trajectory frames, calculate the displacements, velocities and other related properties of each subsystems. Tables~\ref{table:ConstantStressRuns} and \ref{table:ImpulsionsRuns} summarize the characteristics of the trajectories used in the present study. 

The bootstrap analysis~\cite{Press_Flannery_NumericalRecipes_C} was implemented as follows. In each case, a number $N_s$ of realisations $\XX_{\mu}^{(\alpha)}(t)$ of    given procedure (CPF, FKR \ldots with different input parameters) is taken as working sample.
Prior to analysing, a collection of weight vectors $w^{(\alpha)}_{\beta}$; $\beta=1 \ldots M$;  $\alpha=1 \ldots N_s$ was drawn at random, where for each given $\beta$, $N_s$ independent draws of integers $I \in [1,N_s]$ were performed and $w_{\beta}^{(I)}$ was set equal to the number of times $I$ was drawn (with repetition) during the process, and divided by $N_s$. 
In this way $w_{\beta}^{(\alpha)}$ is normalized ( $\sum_{\alpha} w_{\beta}^{(\alpha)}=1$). The flat sample average  corresponds to the special vector $w_{0}{(\alpha)}=1/N_s$. 
Each bootstrap realisation corresponds to a contraction $X_{\beta,\mu}(t) = \sum_{\alpha} w_{\beta}^{(\alpha)} \XX^{(\alpha)}_{\mu}(t)$ of the working sample. Functions $X_{\beta,\mu}(t)$ represent a randomly resampled average of the original working sample, close to the flat average $\mean{\XX_{\mu}} = 1/N_s \sum_{\alpha}\XX^{(\alpha)}_{\mu}(t)$. The relative variation of the quantities of interest deduced from $\mean{\XX_{\mu}}(t)$, such as plateau values or average velocities, provides a confidence interval for the quantity of interest. Bootstrap amounts to randomly selecting subsets of the working sample in order to infer its intrinsic variability. The whole procedure is a kind of Monte-Carlo estimate of an average value, using the working sample as configuration space. For large and independent enough samples, the bootstrap approach should indicate the true variability of the desired average value. Throughout this work, we used twice the square-root deviation $2\sigma_b$ of the bootstrap samples as our confidence interval.

\begin{widetext}
%%%%%%%%%%%%%%%%%%%%%%%%%%%%%%%%%%%%%%%%%%%%%%%%%%%%%%%%%%%%%%%%%%%%%%
%   TABLE OF RUNS
%%%%%%%%%%%%%%%%%%%%%%%%%%%%%%%%%%%%%%%%%%%%%%%%%%%%%%%%%%%%%%%%%%%%%%

%sim_PULL
%%%%%%%%%%%%%%%%%%%%%%%%%%%%%%%%%%%%%%%%%%%%%%%%%%%%%%%%%%%%%%%%%%%%

\begin{table}[]
\begin{tabular}{|c|l|l|c|c|c|}
\hline
\textbf{Simulation type} & 
\multicolumn{1}{c|}{\textbf{State (K)}} & 
\multicolumn{1}{c|}{\textbf{Box size (nm)}} & 
\textbf{Lipids} & 
\textbf{Stress (bar)} & 
\textbf{Number of runs} \\ 
\hline
CPF & fluid (340) & 
\multicolumn{1}{c|}{xy:13.18 ; z:8.17} & 
512-10W & 
\begin{tabular}[c]{@{}c@{}} F=10 ; $\tau = 0.955$ \\ 
F=50 ; $\tau =4.78$ \\ F=100 ; $\tau =9.55$\\ F=150 ; $\tau =14.3$\\ F=200 ; $\tau =19.1$\\ F=250 ; $\tau=23.9$\\ F=500 ; $\tau =47.8$\\ F=1000 ; $\tau =95.5$\\ F=2000 ; $\tau=191$ \\ F=3000 ; $\tau =2.87$ 
\end{tabular} & 
50  \\
\hline
CPF & 
fluid (340) & 
xy:18.55; z:8.24 & 
1024-10W & 
\begin{tabular}[c]{@{}c@{}} 
  F=200; $\tau = 9.66$\\ 
  F=250; $\tau =12.1$\\ 
  F=400; $\tau =19.3$ 
\end{tabular}  & 
50  \\ 
\hline
CPF & 
fluid (340) & 
xy:18.55; z:8.24 & 
1024-10W & 
\begin{tabular}[c]{@{}c@{}} 
  F=200; $\tau = 9.66$\\ 
  F=250; $\tau =12.1$\\ 
  F=400; $\tau =19.3$ 
\end{tabular}  & 
50  \\ 
\hline
CPF  & gel (280) & xy:11.14; z:10.6 & 
\multicolumn{1}{l|}{512-10W} & 
\begin{tabular}[c]{@{}c@{}} 
  F=10 ; $\tau = 1.34$\\ 
  F=50 ; $\tau =6.69$ \\ 
  F=100 ; $\tau =13.4$ \\ 
  F=150 ; $\tau =20.1$ \\ 
  F=200 ; $\tau =26.8$ \\ 
  F=250 ; $\tau =33.4$ \\ 
  F=500 ; $\tau =66.9$ \\ 
  F=1000 ; $\tau =134.$\\ 
  F=2000 ; $\tau =268$\\ 
  F=3000 ; $\tau =401.$ 
\end{tabular}   & 
50  \\
\hline
\end{tabular}
\caption{List of simulations used in the constant pull force statistics.}
\label{table:ConstantStressRuns}
\end{table}

%SIM_IMPULSION
\begin{table}[]

\begin{tabular}{|c|c|c|c|l|c|}
\hline
\textbf{Simulation type} & 
\textbf{State (K)}  & 
\textbf{Box size (nm)} & 
\textbf{nbr Lipids} & 
\multicolumn{1}{c|}{\textbf{Velocity (nm.ps$^{-1}$)}}  & 
\textbf{nbr Runs} \\
\hline
% Second line is special
\multirow{3}{*}{\textbf{FKR}} & 
\multirow{5}{*}{fluid (340)} &
   x-y: 13.18; z: 8.17  & 
   512-10W  & 
    \begin{tabular}[c]{@{}l@{}} 
      0.01:0.1 step=0.01\\ 
      0.1-0.5 step=0.1
    \end{tabular} & 
\multirow{3}{*}{150}  \\
% Third line, only fill columns 3-5
\cline{3-5} &  
   & 
   x-y: 13.15 ; z: 11.88 & 
   512-20W  & 
   \begin{tabular}[c]{@{}l@{}}
     0.05:0.1 step 0.01\\ 
     0.2
   \end{tabular} & 
   \\
% Fourth line, only fill columns 3-5
\cline{3-5} &  
   & 
   x-y: 18.55 ; z: 8.24  & 
   \multicolumn{1}{c|}{1024-10W} & 
   \begin{tabular}[c]{@{}l@{}}
     0.04:0.1 step=0.01\\
     0.2
   \end{tabular} &  
   \\
 % Fifth line
\cline{1-1} \cline{3-6}  \textbf{FKR'}  & 
  & 
  x-y: 13.15 ; z: 8.21  & 
  \multicolumn{1}{c|}{512-10W}  & 
  \begin{tabular}[c]{@{}l@{}}
    0.05-0.10\\
    0.20
  \end{tabular}  & 
    1000  \\ 
  \hline
  % Sixth line
\cline{1-1} \cline{3-6}  \textbf{FKR-Tilt}  & 
  & 
  x-y: 13.18 ; z: 8.17  & 
  \multicolumn{1}{c|}{512-10W}  & 
  \begin{tabular}[c]{@{}l@{}}
    0.09\\
    0.00
  \end{tabular}  & 
    500  \\ 
  \hline
%%%  First Gel line
\textbf{FKR} & 
\multicolumn{1}{l|}{\multirow{3}{*}{gel (280)}} & 
  x-y: 11.14 ; z: 10.64 & 
  \multicolumn{1}{c|}{512-10W}  & 
  \begin{tabular}[c]{@{}l@{}}
    0.01:0.1 step=0.01\\
    0.1-0.5 step=0.1
  \end{tabular} & 
  150 \\  
  % Second gel line
  \cline{1-1} \cline{3-6} 
  \textbf{FKR'} & 
  \multicolumn{1}{c|}{} & 
     x-y: 11.14 ; z: 10.63 & 
     512-10W  & 
   \begin{tabular}[c]{@{}l@{}}
     0.05-0.10\\ 
     0.2,0.3
   \end{tabular}                          
   & 1000 \\ 
   % Third gel line
  \cline{1-1} \cline{3-6} 
  \textbf{FKR-Tilt} & 
  \multicolumn{1}{c|}{} & 
     x-y: 11.14 ; z: 10.64 & 
     512-10W  & 
   \begin{tabular}[c]{@{}l@{}}
     0.09\\ 
     0.00
   \end{tabular}                          
   & 500 \\ 
   \hline
\end{tabular}
\caption{List of simulations used for the force kick relaxation statistics. The notation 512-10W stands for 512 lipids and 10 water beads per lipid.}
\label{table:ImpulsionsRuns}
\end{table}

\end{widetext}

\newpage

%%%%%%%%%%%%%%%%%%%%%%%%%%%%%%%%%%%%%%%%%%%%%%%%%%%%%%%%%%%%%%%%%%%%%%%%%%
\section*{Bibliography}
\bibliography{biblio_sliding}

\newpage

\end{document}

\title{Rheology of sliding leaflets in coarse-grained DSPC lipid bilayers. Supplementary Information (SI)}
\author{Othmene Benazieb}
\affiliation{Institut Charles Sadron, CNRS and University of Strasbourg, 23 rue du Loess, F-67034 Strasbourg cedex 2, France}
\author{Claire Loison}
\affiliation{Institut Lumière Matière, UMR5306 Université Lyon 1-CNRS, Université de Lyon,
69622 Villeurbanne cedex, France}
\author{Fabrice Thalmann}
\affiliation{Institut Charles Sadron, CNRS and University of Strasbourg, 23 rue du Loess, F-67034 Strasbourg cedex 2, France}
\date{\today}
\begin{abstract}
\end{abstract}

\maketitle
\renewcommand{\thefigure}{S\arabic{figure}}

\section*{Supplementary figure}

%%%%%%%%%%%%%%%%%%%%%%%%%%%
\begin{figure}[ht]
\centering
\includegraphics[width=0.7\textwidth,angle=0]{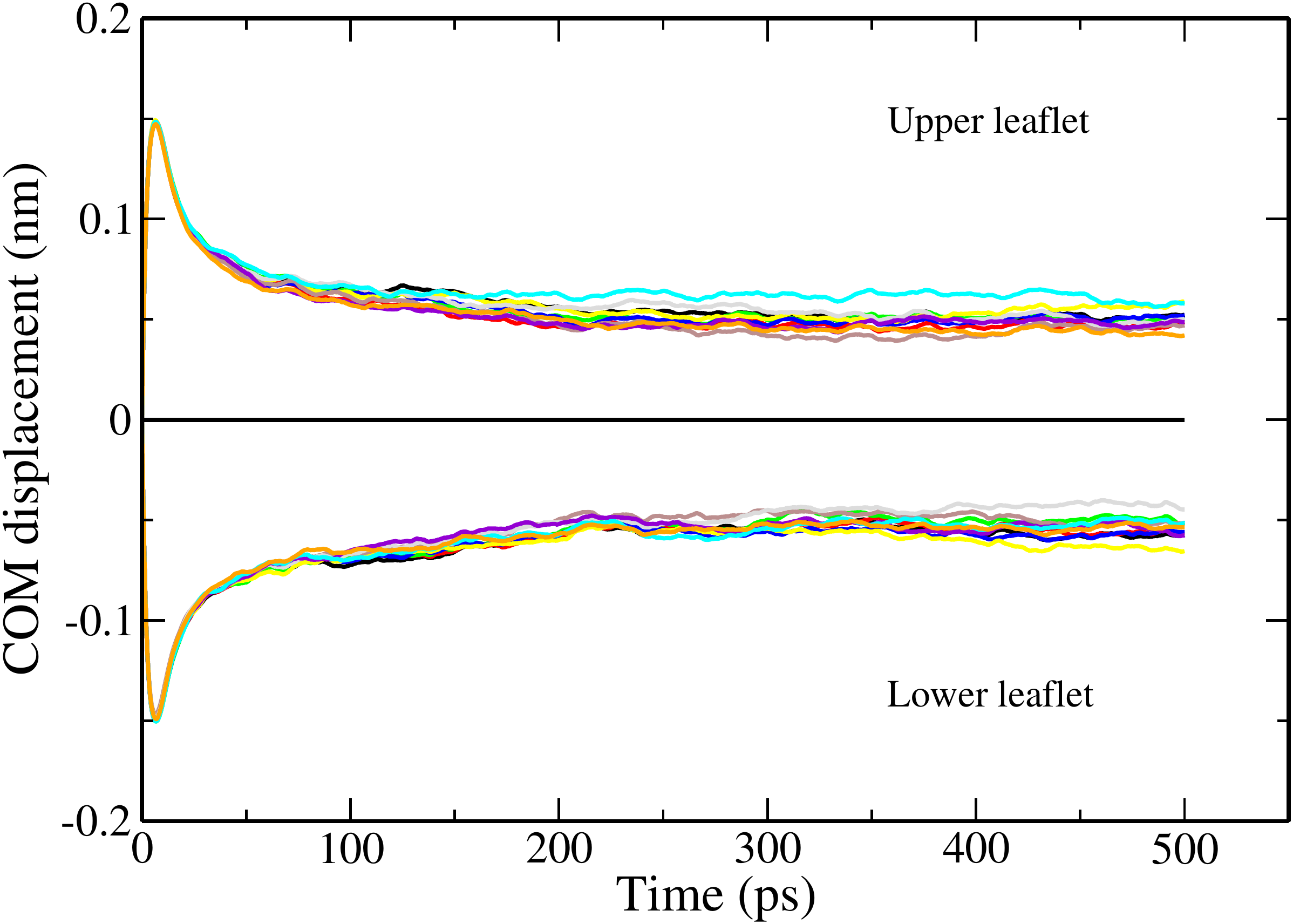}
\caption{Ten different bootstrap realizations of the displacement $\mean{\XX_u}(t)$ for an initial kick impulsion $V_0=0.09~\mathrm{nm.ps}^{-1}$.}
\label{fig:FigS1_bootstrap_displacement_fkr_fluid}
\end{figure}
%%%%%%%%%%%%%%%%%%%%%%%%%%%

%%%%%%%%%%%%%%%%%%%%%%%%%%%
\begin{figure}[ht]
\centering
\includegraphics[width=0.7\textwidth,angle=0]{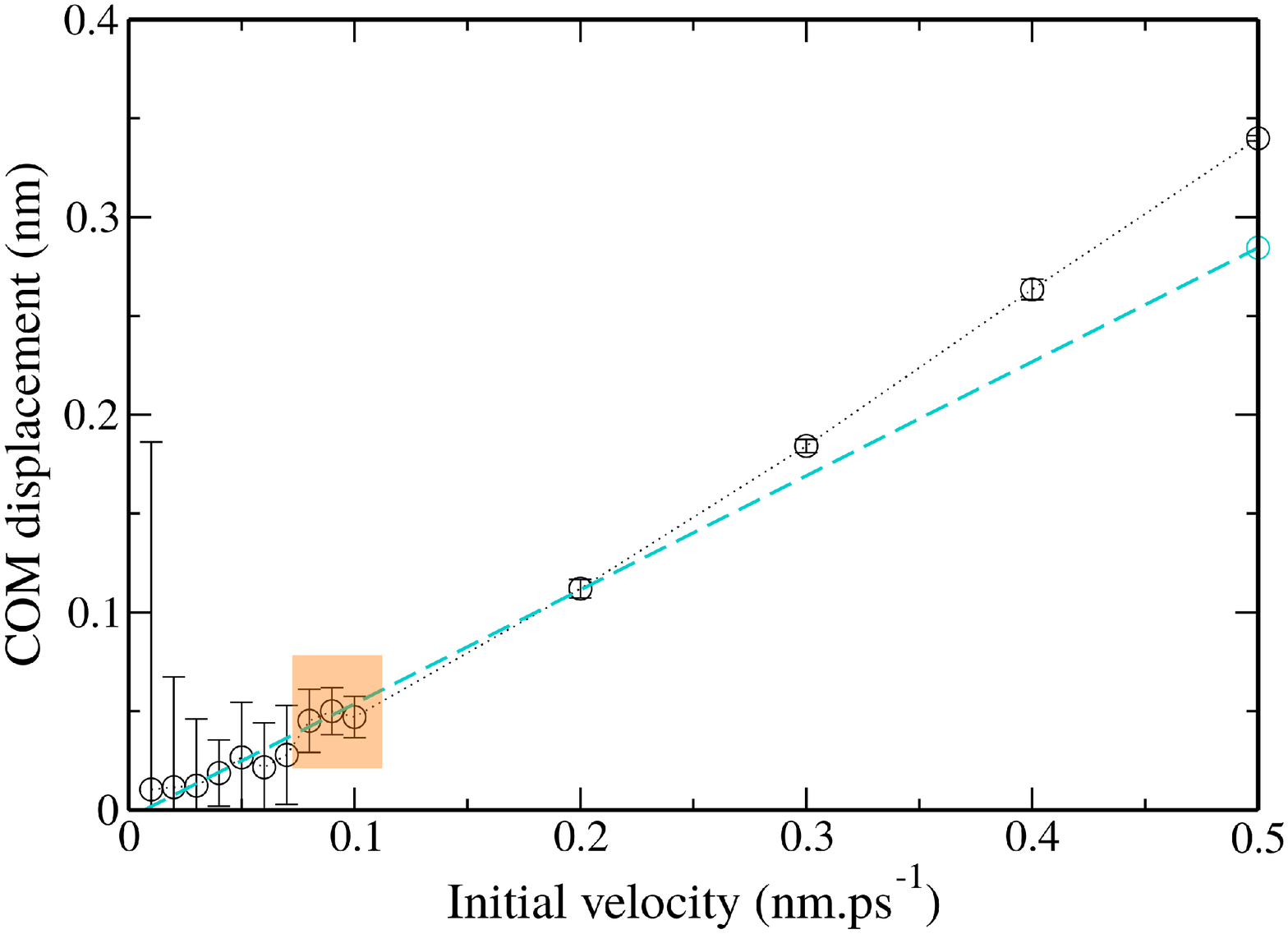}
\caption{Displacement $\mean{\Delta \XX_u}$ as a function of initial velocity increment $V_0$ in the Couette force kick relaxation of a fluid bilayer.}
\label{fig:FigS2_linear_regime_fkr}
\end{figure}
%%%%%%%%%%%%%%%%%%%%%%%%%%%

%%%%%%%%%%%%%%%%%%%%%%%%%%%
%%% Bootstrap displacements in the gel phase
%%%%%%%%%%%%%%%%%%%%%%%%%%%
\begin{figure}[ht]
\centering
\includegraphics[width=0.7\textwidth,angle=0]{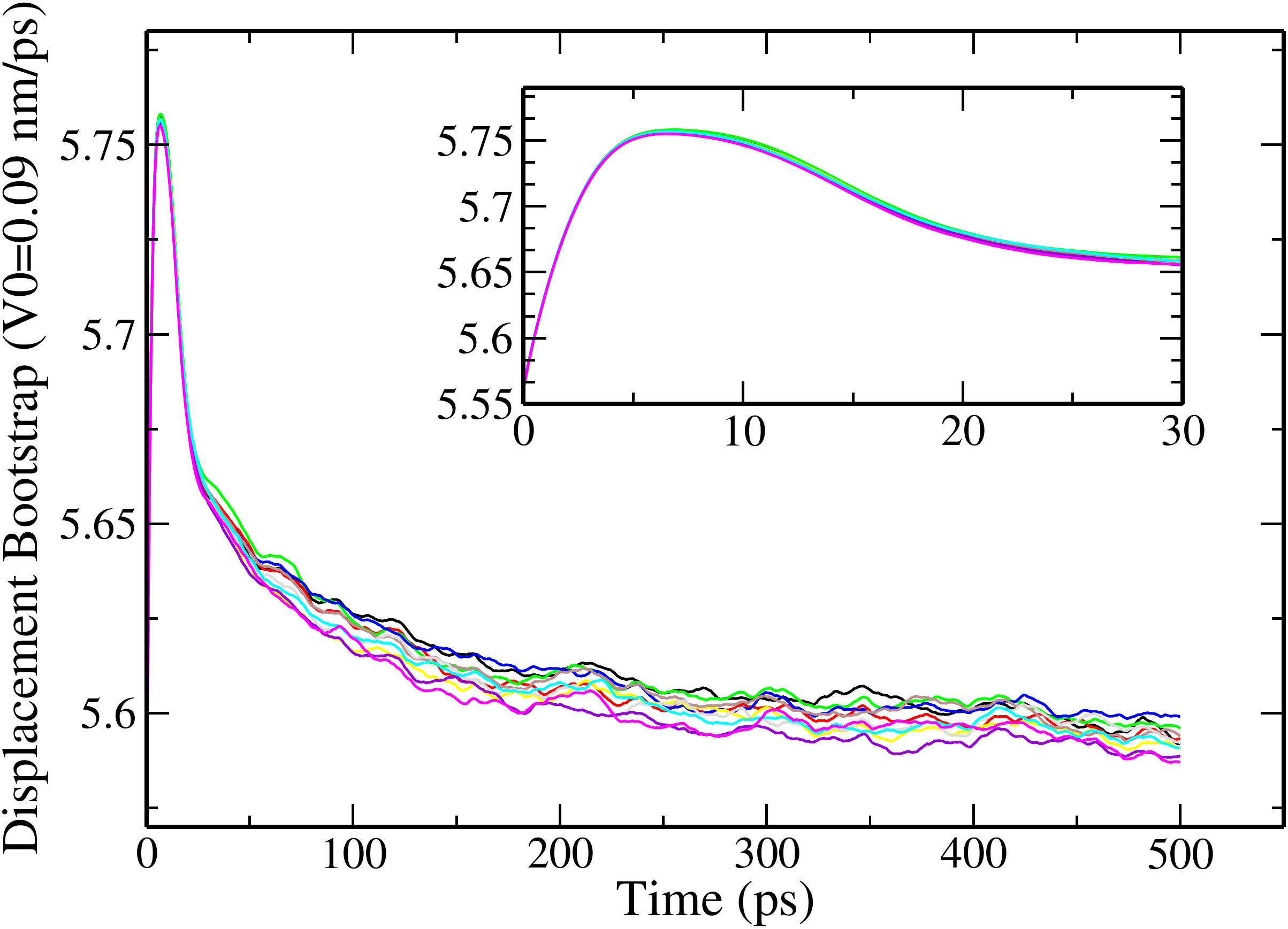}
\caption{Ten different bootstrap realizations of the displacement $\mean{\XX_u}(t)$ for an initial kick impulsion $V_0=0.09~\mathrm{nm.ps}^{-1}$  in the gel phase (upper leaflet).  Inset: close-up look at the first peak.}
\label{fig:FigS3_variability_from_bootstrap_cpf_gel}
\end{figure}

%%%%%%%%%%%%%%%%%%%%%%%%%%%%%%%%%%%%%%%%%%%%%%%%%%%%%%%%%%
%%% Velocity stress characteristics 
%%%%%%%%%%%%%%%%%%%%%%%%%%%%%%%%%%%%%%%%%%%%%%%%%%%%%%%%%%
\begin{figure}[ht]
\centering
\includegraphics[width=0.7\textwidth,angle=0]{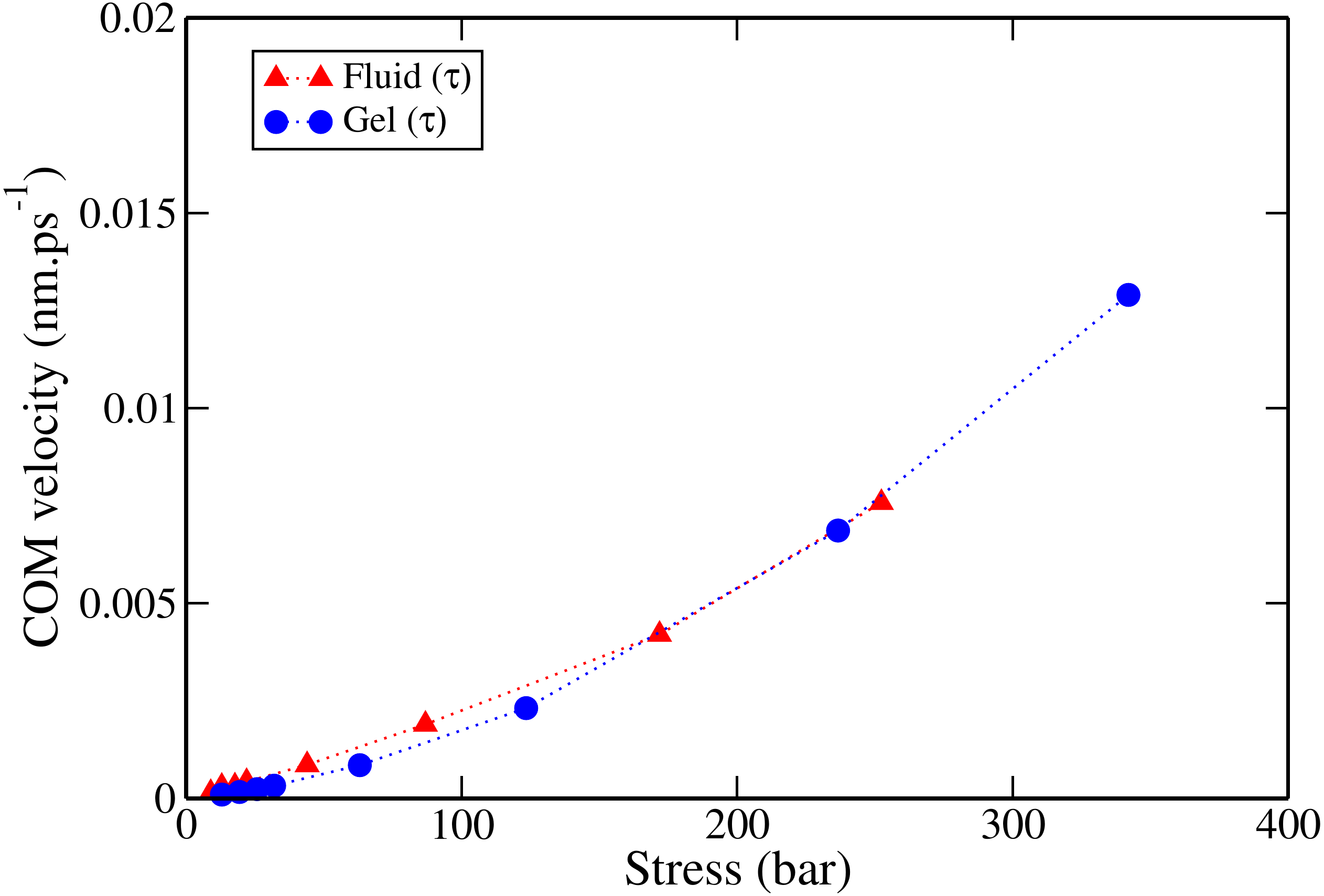}
\caption{Constant pull force (CPF) velocity $V_u$- applied shear stress $\phi_u$ and inner stress $\tau$ characteristics in the gel and fluid states (eq.~27). If the fluid state display a linear characteristics up to $\tau$=100 bars (linear regime) the characteristics seems nowhere linear in the gel state.}
\label{fig:FigS16_gel_and_fluid_vs_stress}
\end{figure}

%%%%%%%%%%%%%%%%%%%%%%%%%%%%%%%%%%%%%%%
%%% Reference Tilt angle
%%%%%%%%%%%%%%%%%%%%%%%%%%%%%%%%%%%%%%%
\begin{figure}[ht]
\centering
\includegraphics[width=0.7\textwidth,angle=0]{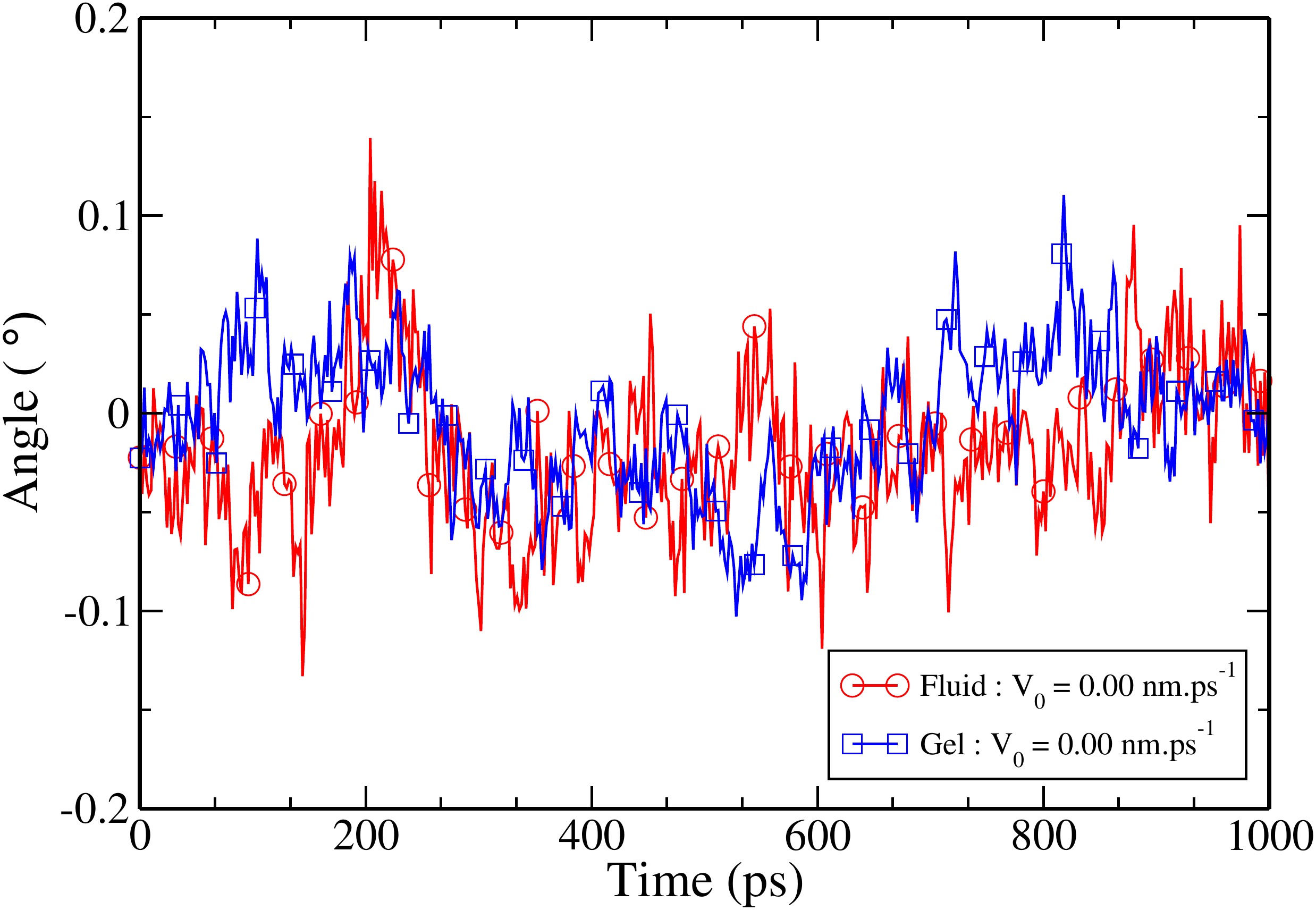}
\caption{Reference tilt angle $\mean{\theta}$ at equilibrium, averaged over 1000 independent samples.}
\label{fig:Fig15_equilibrium_tilt}
\end{figure}